\documentclass[]{aa}
\usepackage{color }
\usepackage{natbib}
\usepackage{amsmath}
\usepackage{amssymb}
\usepackage{graphicx}
\graphicspath{{pictures/}}
\DeclareGraphicsExtensions{.png}
\newcommand{\bl}[1]{\mbox{\boldmath$ #1 $}}
\begin{document}
\title{Gravitoviscous protoplanetary disks with a dust component. I. The importance of the inner sub-au region}
\author{Eduard I. Vorobyov\inst{1,2}, Aleksandr M. Skliarevskii\inst{2}, Vardan G. Elbakyan\inst{2,3}, Yaroslav Pavlyuchenkov\inst{4}, Vitaly Akimkin\inst{4}, and Manuel Guedel\inst{1}}
\institute{ 
University of Vienna, Department of Astrophysics, Vienna, 1180, Austria \\
\email{eduard.vorobiev@univie.ac.at} 
\and
Research Institute of Physics, Southern Federal University, Rostov-on-Don, 344090 Russia
\and 
Lund Observatory, Department of Astronomy and Theoretical Physics, Lund University, Box 43, 22100 Lund, Sweden
\and 
Institute of Astronomy, Russian Academy of Sciences, Pyatnitskaya str. 48, Moscow, 119017, Russia
}

\abstract
{}
{The central region of a circumstellar disk is difficult to resolve in global numerical simulations of collapsing cloud cores, but its effect on the evolution of the entire disk can be significant.}
{We use numerical hydrodynamics simulations to model the long-term evolution of self-gravitating and viscous circumstellar disks in the thin-disk limit. Simulations start from the gravitational collapse of prestellar cores of 0.5--1.0~$M_\odot$ and both gaseous and dusty subsystems were considered, including a model for dust growth. The inner unresolved 1.0 au of the disk is replaced with a central ``smart'' cell (CSC) -- a simplified model that simulates physical processes that may occur in this region. }
{We found that the mass transport rate through the CSC has an appreciable effect on the evolution of the entire disk. Models with slow mass transport form more massive and warmer disks and they are more susceptible to gravitational instability and fragmentation, including a newly identified episodic mode of disk fragmentation in the T Tauri phase of disk evolution. 
Models with slow mass transport through the CSC feature episodic accretion and luminosity bursts in the early evolution, while models with fast transport are characterized by a steadily declining accretion rate with low-amplitude flickering. Dust grows to a larger, decimeter size in the slow transport models and  efficiently drifts in the CSC, where it accumulates reaching the limit when streaming instability becomes operational. We argue that gravitational instability, together with streaming instability likely operating in the inner disk regions, constitute two concurrent planet-forming mechanisms, which may explain the observed diversity of exoplanetary orbits.}
{We conclude that sophisticated models of the inner unresolved disk regions should be used when modeling the formation and evolution of gaseous and dusty protoplanetary disks.}

\authorrunning{Vorobyov et al.}
\titlerunning{Effects of central unresolved disk regions}

\maketitle

\section{Introduction}
Circumstellar disks form during the gravitational collapse of rotating pre-stellar cores and span a range from $\lesssim 0.1$~au, where they are truncated by the stellar magnetosphere, to tens or even hundreds of au.
These vast variations in spatial scales present a challenge both observationally and numerically. Because of high optical depths in the near-infrared, the inner disk regions can 
be best studied using longer wavelength observations, but even the most powerful (sub-)millimeter observational facilities, such as the ALMA or JVLA, cannot resolve the innermost disk regions. The situation is not better with numerical hydrodynamics simulations, as resolving the inner regions of the disk requires sub-au numerical resolution, which for the global disk simulations is prohibitively expensive. Besides, the innermost parts of the disk are also characterized by the highest gas temperature and rotation rate, which imposes strict limitations on the time-explicit numerical schemes. 

To circumvent the problems with the Courant condition, a sink cell of a certain radius is cut around the central star in numerical hydrodynamics simulations. This trick, however, brings about undesirable complications. Firstly, an important physical region of the disk is eliminated from simulations, which otherwise could have influenced the disk evolution via accretion and luminosity outbursts triggered by various instabilities that can occur there (e.g., MRI and/or thermoviscous instability).  \citet{2011Stamatellos} realized this problem and used a toy model of FU-Orionis-type outbursts implemented in the sink cell treatment based on the model of \citet{2009ZhuHartmannGammie}.
Secondly, the interface between the inactive sink and the inner active disk presents a numerical challenge, as it is not trivial to devise the proper boundary conditions, so that a smooth transition in gas and dust density through the sink cell--disk interface can be established. This problem was attacked by \citet{2007Crida}, who employed a simplified one-dimensional model for the sink, and by \citet{2018VorobyovAkimkin}, who devised a sophisticated inflow-outflow boundary condition allowing for matter to flow in both directions through the sink cell--disk interface. In the context of smoothed particle hydrodynamics, \citet{2013Hubber} devised an improved sink particle model, which accreted matter, but allowed for the associated angular momentumto is transferred back to the surrounding medium. Finally, the size of the sink cell can affect the disk formation phase, impeding disk formation due to magnetic braking if the sink radius greatly exceeds one au \citep{2014Machida}.

In this work, we address the effects that the inner unresolved sub-au regions may have on the global long-term evolution of gaseous and dusty circumstellar disks. 
The first study of this kind was performed by \citet{2014Ohtani}, who found that the inner disk can alter the character of mass transport and the resulting mass accretion on the star can be different from the mass transport in the inner several au of the disk.   Unlike \citet{2014Ohtani}, in this study we mostly focus on effects that the physical processes of mass transport in the inner disk may have on the global properties of the entire star plus disk system. 
We develop a simple toy model for the sink cell, which is hereafter referred as the central smart cell (CSC) by the reasons described in Sect.~\ref{sinkmodel}. The developed model is meant to describe the physical processes in the inner 1.0 au of the disk; it reproduces episodic accretion bursts, as expected for young protostars, and allows for a smooth transition in the gas and dust density through the CSC--disk interface. We consider various efficiencies of the mass transport rate through the CSC that may correspond to the presence or absence of the dead zone in the inner 1.0 au of the disk and investigate its effect on the global structure and evolution of a gravitoviscous disk in the limit of a fully developed magnetorotational instability.

The paper is organized as follows. In Sect.~\ref{diskmodel} we review the main component of our numerical hydrodynamics model. In Sect.~\ref{sinkmodel} we present the model for the CSC. Sect.~\ref{BI} reviews the adopted boundary and initial conditions. The main results are discussed in Sect.~\ref{results} and summarized in Sect.~\ref{summary}.

\section{Numerical model}
\label{diskmodel}

Numerical simulations of disk formation and long-term evolution present a computational challenge. They have to capture spatial scales from sub-au to thousands of au with a numerical resolution that is sufficient to resolve disk substructures on au or even sub-au scales (e.g., gravitational fragmentation, streaming instability, etc.). Since most of our knowledge about disk characteristics is derived from observations of T Tauri disks, numerical codes need to follow disk evolution from their formation in the embedded, optically obscured phase to the optically revealed T~Tauri phase. This implies evolution times on the order of a million years \citep{2009EvansDunham}. These requirements make fully three-dimensional simulations prohibitively expensive. 3D models with a sink cell $\ge$~5--10~au and integration times $\le 0.1$~Myr require enormous computational resources -- thousands of CPU cores and millions of CPU hours -- not easily available for the typical institutional infrastructure.

This is the reason why simpler numerical models of disk formation and long-term evolution are still widely used to derive disk characteristics for a wide parameter space. The most often used approach is to employ various modifications of the one-dimensional viscous evolution equation for the surface density of gas as derived by \citet{1981Pringle}
\begin{equation}
{\partial \Sigma_{\rm g} \over \partial t} + {1 \over r} {\partial \over \partial r} { 1 \over 
(r^2 \Omega)^\prime }  {\partial \over \partial r} \left( \mu r^3 \Omega^\prime  \right) =0,
\label{dif1}
\end{equation}
where $\Sigma_{\rm g}$ is the gas surface density, $\Omega$ is the angular velocity, $\mu$ is the dynamic viscosity. and $^\prime$ denotes the radial derivative.
The general drawback of this approach is the inherent lack of azimuthal disk sub-structures, such as spiral arms, gaseous clumps, vortices, etc. These sub-structures can play a substantial role in the global structuring of protostellar/protoplanetary disks, including dust evolution and growth. Moreover, the underlying assumption of negligible gas pressure, which is used to derive Equation~(\ref{dif1}), makes dust evolution simulations within the framework of this model inconsistent. 

These limitations can be lifted in the disk models employing numerical hydrodynamics equations in the thin-disk limit. Much in the same way as the one-dimensional viscous evolution equation~(\ref{dif1}), the thin-disk models assume negligible vertical motions and the local hydrostatic equilibrium. An additional assumption of a small vertical disk scale height with respect to the radial position in the disk (not exceeding 10\%-20\% of the radial distance from the star) if the  allows for the integration of the main hydrodynamic quantities in the vertical direction and the use of these integrated quantities in the hydrodynamics equations.

The obvious advantage of the 2D thin-disk numerical hydrodynamics models is that they, unlike the 1D viscous evolution equation, can self-consistently account for the effects of asymmetric disk sub-structures (spiral arms, clumps, vortices, etc.) on the global evolution of gas and dust in the disk, being at the same time computationally inexpensive in comparison to the full three-dimensional approach. Our code runs on 32 cores (a typical server or a computer node in a cluster) and takes about one week to compute 1.0~Myr of disk evolution.
The disadvantage of the thin-disk models is the lack of the disk vertical structure, although models now start to emerge that relax this limitation  and allow for the on-the-fly reconstruction of the disk vertical structure \citep{2017VorobyovPavlyuchenkov}.  Although intrinsically limited in its ability to model the full set of physical phenomena that may occur in star and planet formation, the thin-disk models nevertheless present an indispensable tool for studying the long-term evolution of protoplanetary disks for a large number of model realizations with a much higher realism than is offered by the simple one-dimensional viscous disk models.

The numerical model for the formation and evolution of a star and its circumstellar disk in the thin-disk limit $(r, \phi)$ is described in detail in \citet{2018VorobyovAkimkin} and is updated to include the  treatment of back reaction of dust on gas. Here, we briefly review its main constituent parts and equations. Numerical simulations start from a collapsing pre-stellar core of a certain mass, angular momentum, temperature, and dust-to-gas mass ratio and terminate after 1.0 Myr of evolution in the T Tauri phase.  The characteristics of the nascent star (radius, photospheric luminosity) are calculated using the stellar evolution tracks obtained with the STELLAR code \citep{2008YorkeBodenheimer}.  The evolution of the star and disk are interconnected. The star accretes matter from the disk 
and feeds back to the disk via radiative heating according to its photospheric and accretion luminosities.

\subsection{FEOSAD code: the gaseous component}
\label{gaseous}

The main physical processes considered in the FEOSAD code when modeling the disk formation and evolution include viscous and shock heating, irradiation by the forming star,  background irradiation with a uniform temperature of $T_\mathrm{bg}=20$\,K 
set equal to the initial temperature of the natal cloud core,
radiative cooling from the disk surface, friction between the gas and dust components, and self-gravity of gaseous and dusty disks.  The code is written in the thin-disk limit, complemented by a calculation of the gas vertical  scale height using an assumption of local hydrostatic equilibrium as described in \citet{VorobyovBasu2009}. The resulting  model has a flared structure (because the disk vertical scale height increases with radius), which guarantees that both the disk and envelope receive a fraction of the irradiation energy  from the central protostar. The pertinent equations of mass, momentum, and energy transport for the gas component are
\begin{equation}
\label{cont}
\frac{{\partial \Sigma_{\rm g} }}{{\partial t}}   + \nabla_p  \cdot 
\left( \Sigma_{\rm g} \bl{v}_p \right) =0,  
\end{equation}
\begin{eqnarray}
\label{mom}
\frac{\partial}{\partial t} \left( \Sigma_{\rm g} \bl{v}_p \right) +  [\nabla \cdot \left( \Sigma_{\rm
g} \bl{v}_p \otimes \bl{v}_p \right)]_p & =&   - \nabla_p {\cal P}  + \Sigma_{\rm g} \, \bl{g}_p + \nonumber
\\ 
&+& (\nabla \cdot \mathbf{\Pi})_p  - \Sigma_{\rm d,gr} \bl{f}_p,
\end{eqnarray}
\begin{equation}
\frac{\partial e}{\partial t} +\nabla_p \cdot \left( e \bl{v}_p \right) = -{\cal P} 
(\nabla_p \cdot \bl{v}_{p}) -\Lambda +\Gamma + 
\left(\nabla \bl{v}\right)_{pp^\prime}:\Pi_{pp^\prime}, 
\label{energ}
\end{equation}
where subscripts $p$ and $p^\prime$ refer to the planar components
$(r,\phi)$  in polar coordinates, $\Sigma_{\rm g}$ is the gas 
surface density,  $e$ is the internal energy per surface area,  ${\cal P}$
is the vertically integrated gas pressure calculated via the ideal  equation of state as ${\cal P}=(\gamma-1) e$ with $\gamma=7/5$, $\bl{v}_{p}=v_r
\hat{\bl r}+ v_\phi \hat{\bl \phi}$  is the gas velocity in the disk plane, and $\nabla_p=\hat{\bl r} \partial / \partial r + \hat{\bl
\phi} r^{-1} \partial / \partial \phi $ is the gradient along the planar coordinates of the disk. The term $\bl{f}_p$ is the drag force per unit mass between dust and gas. A similar term enters the equation of dust dynamics, meaning that we take the back reaction of dust on gas into account.

The gravitational acceleration in the disk
plane,  $\bl{g}_{p}=g_r \hat{\bl r} +g_\phi \hat{\bl \phi}$, takes into account self-gravity of the gaseous and dusty disk components found by solving for the Poisson integral \citep[see details in][]{2010VorobyovBasu} and the
gravity of the central protostar when formed. Turbulent viscosity is
taken into account via the viscous stress tensor  $\mathbf{\Pi}$, the expression for which can be found in \citet{2010VorobyovBasu}. We parameterized the
magnitude of kinematic viscosity $\nu=\alpha c_{\rm s} H_{\rm g}$  using the $\alpha$-prescription of \citet{1973ShakuraSunyaev} with a constant
$\alpha$-parameter set equal to 0.01, where $c_{\rm s}$ is the sound speed of gas and $H_{\rm g}$ is the gas vertical scale height. The expressions for the cooling and heating rates $\Lambda$ and $\Gamma$
can be found in \citet{2018VorobyovAkimkin}.

\subsection{FEOSAD code: the dusty component}
\label{dustycomp}
In our model, dust consists of two components: small micron-sized dust and grown dust.  The former constitutes the initial reservoir
for dust mass and is gradually converted in grown dust as the disk forms and evolves.   Small dust is assumed to be coupled to gas, meaning that we only solve the continuity equation for small dust grains, while the dynamics of grown dust is controlled by friction with the gas component and by the total gravitational potential of the star, gaseous and dusty components. The conversion of small to grown dust is considered by calculating the dust growth rate $S$.
The resulting continuity and momentum equations for small and grown dust are
\begin{equation}
\label{contDsmall}
\frac{{\partial \Sigma_{\rm d,sm} }}{{\partial t}}  + \nabla_p  \cdot 
\left( \Sigma_{\rm d,sm} \bl{v}_p \right) = - S(a_{\rm r}),  
\end{equation}
\begin{equation}
\label{contDlarge}
\frac{{\partial \Sigma_{\rm d,gr} }}{{\partial t}}  + \nabla_p  \cdot 
\left( \Sigma_{\rm d,gr} \bl{u}_p \right) = S(a_{\rm r}),  
\end{equation}
\begin{eqnarray}
\label{momDlarge}
\frac{\partial}{\partial t} \left( \Sigma_{\rm d,gr} \bl{u}_p \right) +  [\nabla \cdot \left( \Sigma_{\rm
d,gr} \bl{u}_p \otimes \bl{u}_p \right)]_p  &=&   \Sigma_{\rm d,gr} \, \bl{g}_p + \nonumber \\
 + \Sigma_{\rm d,gr} \bl{f}_p + S(a_{\rm r}) \bl{v}_p,
\end{eqnarray}
where $\Sigma_{\rm d,sm}$ and $\Sigma_{\rm d,gr}$ are the surface
densities of small and grown dust, $\bl{u}_p$ describes the planar components of the grown dust velocity, $S(a_{\rm r})$ is the rate of small-to grown dust conversion per unit surface area, and $a_{\rm r}$ is the maximum radius of grown dust.
In this study, we assume that dust and gas are vertically mixed, which is justified in young gravitationally unstable and/or MRI-active disks \citep{2004RiceLodato,2018Yang}.
However, in more evolved disks dust settling becomes significant. Its effect on the global evolution of gaseous and dusty disks was considered in our previous study \citep{2018VorobyovAkimkin}, showing that dust settling accelerates somewhat the small-to-grown dust conversion.

Equations~(\ref{cont})--(\ref{energ}) and 
(\ref{contDsmall})--(\ref{momDlarge}) are solved using
the operator-split solution procedure consisting of the transport and source steps. In the transport
step, the update of hydrodynamic quantities due to advection
is done using the third-order piecewise parabolic interpolation
scheme of \citet{1984Colella}. In the source step, the update of hydrodynamic quantities
due to gravity, viscosity, cooling and heating, and also
friction between gas and dust components is performed.
This step also considers the transformation of small to
grown dust and also the increase in dust radius $a_{\rm r}$ due
to growth. The small-to-grown dust conversion is defined as
\begin{equation}
\label{GrowthRate}
S(a_{\rm r}) = - {1 \over \Delta t } \Sigma_{\rm d,tot}^n  
{ \int \limits_{a_{\rm r}^n} \limits^{a_{\rm r}^{n+1}} a^{3-p} da \int \limits_{a_{\rm min}} 
\limits^{a_\ast} a^{3-p} da \over \int \limits_{a_{\rm min}} \limits^{a_{\rm r}^n} a^{3-p} da 
\int \limits_{a_{\rm min}} \limits^{a_{\rm r}^{n+1}} a^{3-p} da   },
\end{equation}
where  $\Sigma_{\rm d,tot}=\Sigma_{\rm d,gr}+
\Sigma_{\rm d,sm}$ is the total surface density of dust,  indices $a_{\rm r}^n$ and $a_{\rm
r}^{n+1}$ the maximum dust radii at the current and next time
steps, $a_{\rm min}=0.005~\mu m$ the minimum radius of small dust grains,
$a_\ast=1.0$~$\mu$m a threshold value between small and grown
dust components, and $p=3.5$ the slope of dust distribution over radius.
The evolution of the maximum radius $a_{\rm r}$ is described as:
\begin{equation}
{\partial a_{\rm r} \over \partial t} + (u_{\rm p} \cdot \nabla_p ) a_{\rm r} = \cal{D},
\label{dustA}
\end{equation}
where the growth rate $\cal{D}$ accounts for the dust evolution due to
coagulation and fragmentation. More details on the dust growth scheme can be found in \citet{2018VorobyovAkimkin}. 

We use the polar coordinates ($r,\phi$) on a two-dimensional
numerical grid with $256\times256$ grid zones. The radial grid
is logarithmically spaced, while the azimuthal grid is equispaced.
To avoid too small time steps, we introduce a central axisymmetric cell 
with a radius of  1.0~au (see for its detailed description below). The use of the logarithmically spaced grid in the $r$-direction and 
equidistant grid in the $\phi$-direction allowed us to resolve the disk in the vicinity of the central cell with a numerical resolution as small as 0.036~au.

\section{Model of the central smart cell}
\label{sinkmodel}

\begin{figure*}
\begin{centering}
\includegraphics[scale=0.3]{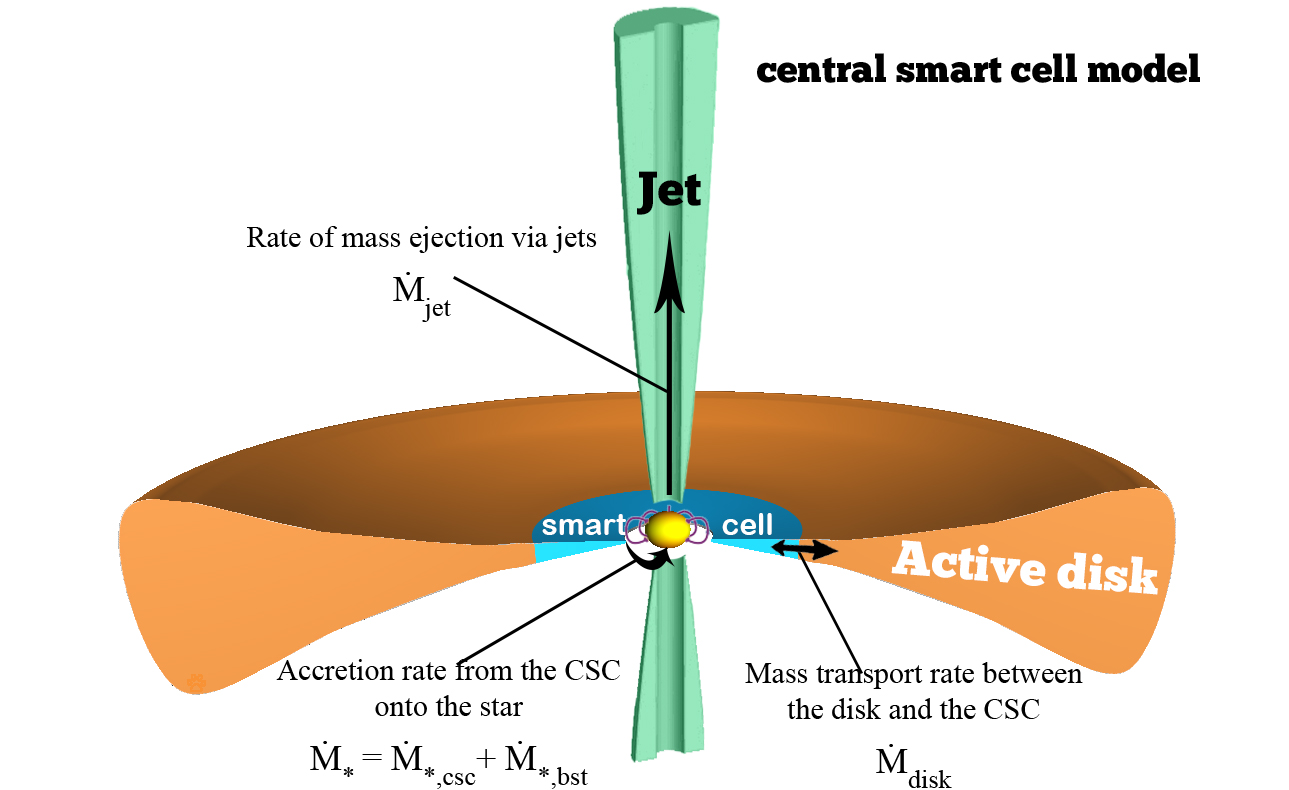}
\par \end{centering}
\caption{Graphical representation of the central smart cell (CSC) in the general context of the protostellar disk model. The main mass transport rates through the CSC are indicated with the arrows.}
\label{CSCM}
\end{figure*}
In this section, we describe the adopted model for the central
cell. Since the model allows for matter to flow in both directions through the interface between the central cell and the active disk, the term "sink cell" (often used in numerical simulations of protoplanetary disks) is no more appropriate. In the following text, we use the term "central smart cell" (hereafter CSC), emphasizing that the central cell has a sub-grid physics, which we describe below.

Our model of the CSC calculates the total mass of gas and dust that is contained in the innermost unresolved disk region, taking in a simplified manner the MRI into account.
The radius of this region is set to $r_{\rm csc}$=1.0~au. The balance of mass in the CSC is computed using the following system of ordinary differential equations
\begin{equation}
{dM_{\rm csc} \over{dt} }= \dot{M}_{\rm disk}-\dot{M}_{*, \rm csc}-\dot{M}_{*, \rm bst}-\dot{M}_{\rm jet}
\label{csc1}
\end{equation}
\begin{equation}
\label{1}
{dM_{*} \over{dt} } = \dot{M}_{*, \rm csc}+\dot{M}_{*,\rm  bst}
\end{equation}
Here, $\dot{M}_{\rm disk}$ is the mass accretion rate through the CSC--disk interface calculated in the hydrodynamics code as the mass flux passing through the inner disk boundary. We note that $\dot{M}_{\rm disk}$ can be both positive (the matter  flows from the disk to the CSC) and negative (the matter flows from the CSC to the disk), depending on the direction of velocity vector at the CSC--disk interface. The mass accretion rate from the CSC on the star is split into two parts: $\dot{M}_{*, \rm csc}$ reflecting the regular mode of mass accretion from the CSC to the star  and $\dot{M}_{*, \rm bst}$  denoting the burst mode of accretion caused by the thermally ignited MRI in the CSC. The former is calculated as
\begin{equation}
\dot{M}_{*, \rm csc}=
\begin{cases}
\xi \dot{M}_{\rm disk} &\text{for $\dot{M}_{\rm disk} > $ 0}, 
\label{Sink:1}
\\ 
0 &\text{for $\dot{M}_{\rm disk} \leqslant $ 0},
\end{cases}
\end{equation}
where we assumed that mass accretion from the CSC on the star is a fraction $\xi$ of mass accretion from the disk to the CSC. The limit of small $\xi  \approx 0$ corresponds to slow mass transport through the CSC, resulting in gradual mass accumulation in the sink. Slow mass transport may be caused by the development of a dead zone in the CSC. The opposite limit of large $\xi \approx 1.0$ assumes fast mass transport, so that the matter that crosses the CSC--disk interface does not accumulate in the CSC, but is quickly delivered on the star. Physically this means that mass transport mechanisms of similar efficiency operate in the disk and in the CSC.

The $\dot{M}_{*, \rm bst}$ term in Equation~(\ref{csc1}) accounts for accretion bursts, which may occur in the CSC through the thermal ignition of the MRI. The MRI burst activates when the gas temperature exceeds the thermal ionization threshold, usually assumed to be above 1000~K. Since we do not calculate the gas temperature in the CSC, we instead assume that the burst is activated when the gas surface density in the CSC exceeds $2\times 10^4$~g~cm$^{-2}$, in accordance with the results of numerical hydrodynamics modeling of \citet{2014BaeHartmann}. We check the gas temperature at the CSC--disk interface at the onset of the burst and confirm that the gas temperature exceeds 1000~K.  When the MRI burst activates, the accretion rate is set equal to $\dot{M}_{*,\rm bst} = 10^{-4}~M_{\odot}$~yr$^{-1}$. If the burst duration exceeds 70~yr or the gas surface density in the central cell drops below 100~g~cm$^{-2}$, the burst terminates. Of course, the adopted parameters of the bursts can vary within a factor of several, but this is not expected to affect our simulations qualitatively.

The last term on the right hand side of Equation~(\ref{csc1}) is the mass ejection rate by the jets. It is defined as $\dot{M}_{\rm jet}=\eta(\dot{M}_{*,\rm csc}+\dot{M}_{*,\rm bst})$, where $\eta$ is a portion of matter that is ejected by the jets. The actual value of $\eta$ varies from 0.01 to 0.5. In this research we use a fixed value of $\eta=0.1$ and leave a more detailed investigation of the effects of varying $\eta$ for a further study. Finally, we note that Equations~(\ref{csc1}) and (\ref{1}) should be applied separately to the gas and dust components. The system of Equations~(\ref{csc1}) and (\ref{1}) is solved at each hydrodynamical time step to obtain the masses of gas and dust in the CSC.
The graphical representation of the CSC model is given in Figure~\ref{CSCM}.

\section{Boundary and initial conditions}
\label{BI}

The choice of boundary conditions in core collapse simulations
always presents a certain challenge. While the outer
boundary can be reflecting, meaning that no matter can
cross the boundary, the inner boundary cannot be of that
type. The code must allow for matter to freely flow through the CSC-disk interface. If the inner boundary allows for matter to flow
only in one direction, i.e., from the disk to the CSC,
then any wave-like motions near the inner computational boundary, such
as those triggered by spiral density waves in the disk, would
result in a disproportionate flow through the CSC-disk interface.
As a consequence, an artificial depression in the gas density
near the inner boundary develops in the course of time because
of the lack of compensating back flow from the CSC
to the disk. This was the reason why $\dot{M}_{\rm disk}$ in Equation~(\ref{csc1}) can have both the positive or negative sign depending on the direction of gas/dust flow through the CSC-disk interface. 

To proceed with numerical simulations, we need to define the values of surface densities, velocities, and gas internal energy at the computational boundaries. Let us first consider the inner computational boundary. 
The surface densities of gas and dust at the inner boundary can be directly obtained by solving Equations~(\ref{csc1}) and (\ref{1}). We note that Equation~(\ref{csc1}) returns the masses of gas and dust in the CSC, but these quantities can be turned into the corresponding surface densities when divided by the fixed area of the CSC. We do not have information on the internal energy or velocity in the CSC. Therefore, the radial velocity and internal energy at the inner boundary are determined from the zero gradient condition at the CSC-disk interface, while the azimuthal velocity is extrapolated from the active disk to the CSC assuming a Keplerian rotation. We emphasize that these boundary conditions enable a smooth transition of the surface density between the inner active disk and the CSC, preventing (or greatly reducing) the formation of an artificial drop in the surface density near the inner boundary. A more detailed study of the effects of the inner boundary condition on the disk evolution is deferred for a separate study. We ensure that our boundary conditions conserve the total mass budget of gas and dust in the system. Finally, we note that the outer boundary condition (usually located at several thousands of au from the star) is set to a standard free outflow, allowing for material to flow out of the computational domain, but not allowing any material to flow in. The zero-gradient condition is applied to all hydrodynamic quantities.

The initial radial profile of the gas surface density $\Sigma_{\rm g}$ and
angular velocity $\Omega$ of the pre-stellar core has the
following form: 
\begin{equation}
\Sigma_{\rm g}=\frac{r_{0}\Sigma_{\rm g,0}}{\sqrt{r^{2}+r_{0}^{2}}},
\label{eq:sigma}
\end{equation}
\begin{equation}
\Omega=2\Omega_{0}\left(\frac{r_{0}}{r}\right)^{2}\left[\sqrt{1+\left(\frac{r}{r_{0}}\right)^{2}}-1\right],
\label{eq:omega}
\end{equation}
where $\Sigma_{\rm g,0}$ and $\Omega_{0}$ are the angular velocity and
gas surface density at the center of the core, $r_{0}=\sqrt{A}c_{\mathrm{s}}^{2}/\pi G\Sigma_{\rm g,0}$
is the radius of the central plateau, where $c_{\mathrm{s}}$ is the initial isothermal sound speed in the core. This radial profile is typical of pre-stellar cores formed as a result of the slow expulsion of magnetic field due to ambipolar diffusion, with the angular momentum remaining constant during axially-symmetric core compression \citep{1997Basu}. The value of the positive density 
perturbation \textit{A} is set to 1.1, making the core unstable to collapse. The initial gas temperature in collapsing cores is $T_{\mathrm{init}}=20\,\mathrm{K}$. We consider a numerical model with $\Omega_{0}=2.05$~km~s$^{-1}$~pc$^{-1}$, 
$\Sigma_{\rm g,0}=0.2~\mathrm{g\,cm^{-2}}$, and $r_{0}=1200\,\mathrm{AU}$.
The resulting core mass $M_{\rm core}=1.03\,M_{\odot}$ and
the ratio of rotational to gravitational energy $\beta_{\rm rot}=2.4\times10^{-3}$. The initial dust-to-gas ratio is 1:100. All dust initially is in the form of small dust particles with a radius of 1.0~$\mu$m. 

\section{Results}
\label{results}

\begin{figure*}
\begin{centering}
\includegraphics[scale=0.2]{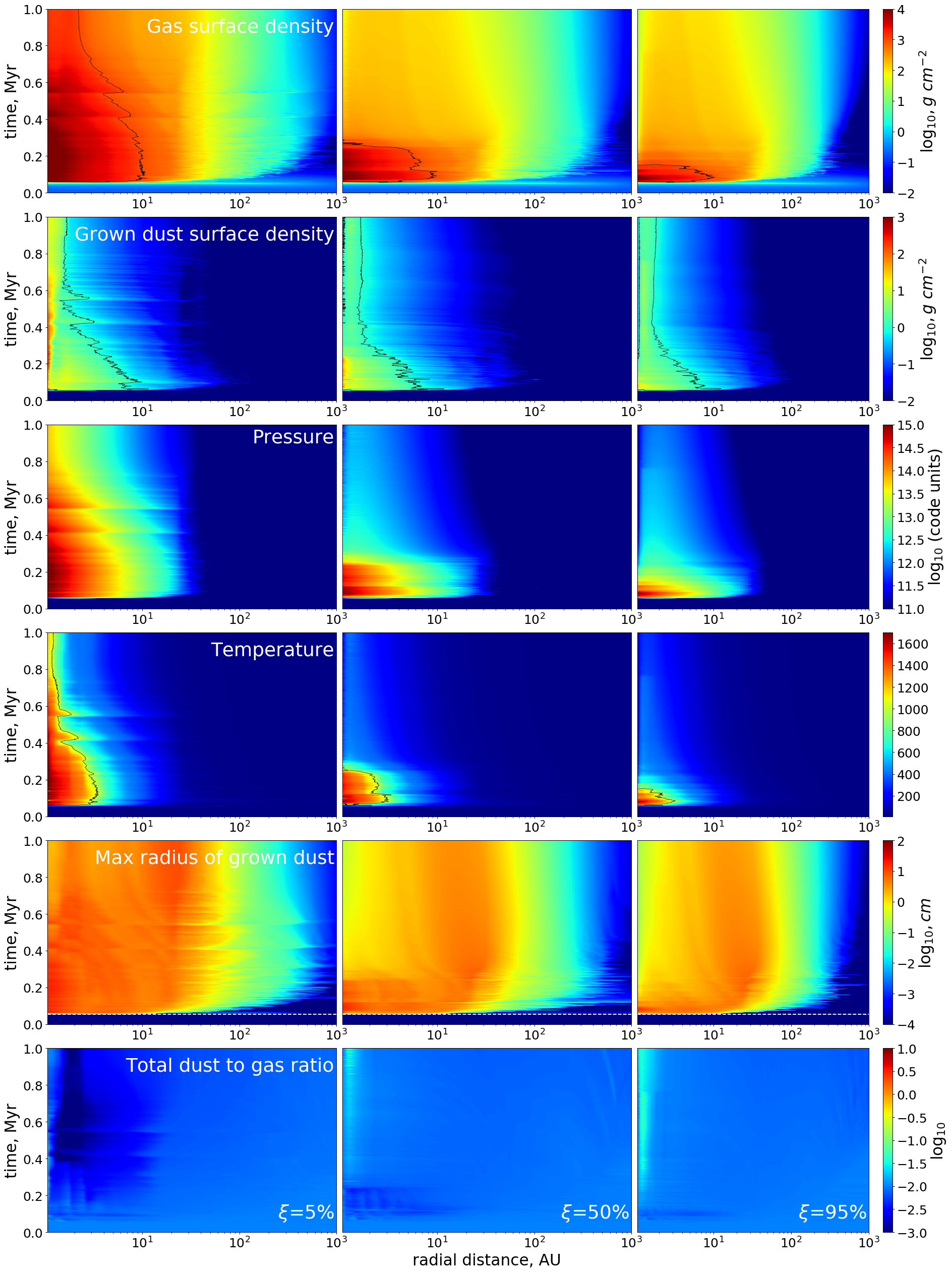} 
\par \end{centering}
\caption{ Temporal evolution of the azimuthally-averaged gas surface density (top row), grown dust surface density (second row), gas pressure (third row), gas temperature (fourth row), maximum radius of grown dust (fifth row), and the dust-to-gas mass ratio (bottom row) in models with $\xi$ = 5\% (left column), $\xi$=50\% (middle column), and $\xi$=95\%  (right column). Contour lines indicate gas surface densities of $10^3$~g~cm$^{-2}$ (top row), grown dust surface densities of  10~g~cm$^{-2}$ (second row), and gas temperatures of  1000~K to emphasize evolutionary trends. The horizontal dotted lines in the fifth row mark the disk formation instance. } 
\label{fig:1}
\end{figure*}

In this work, we consider three values of the $\xi$-parameter controlling the rate of mass transport through the CSC. More specifically, $\xi$ = 0.05 corresponds to slow mass transport through the CSC when only 5\% of the mass that passes through the cell-disk interface (per unit time) arrives at the star and the remaining 95\% is retained in the CSC. The choice of $\xi=0.95$ corresponds to the fast transport regime when 95\% of the mass accretes at the star and only 5\% is retained in the CSC, $\xi$=0.5  corresponds to the intermediate case.

In our model, the mass transport rate in the disk is controlled by viscous and gravitational torques. While gravitational torques operate self-consistently in our numerical simulations thanks to computed disk self-gravity, viscous torques are parameterized using the $\alpha$-prescription of \citet{1973ShakuraSunyaev}. 
The viscous $\alpha$-parameter in the disk is set equal to 0.01, which is characteristic of a fully operational MRI \citep[e.g.,][]{2018Yang}. Lower values of the $\alpha$-parameter will be considered in a follow-up study. The starting point of each simulation is the gravitational collapse of a prestellar core.
In the adopted thin-disk approximation, the core has the form of a flattened pseudo-disk, a spatial configuration that can be expected in the presence of rotation and large-scale magnetic fields \citep[e.g.,][]{1997Basu}. The initial ratio of rotational-to-gravitational energy is only $\beta_{\rm rot}=2.4\times 10^{-3}$ and the core is initially unstable to gravitational collapse. As the collapse proceeds, the inner regions of the core spin up and a  centrifugally balanced circumstellar disk forms at $t=55$~kyr when the inner infalling layers of the core hit the centrifugal barrier near the CSC. The material that has passed to the CSC before the instance of circumstellar disk formation constitutes a seed for the central star, which further grows through accretion from the circumstellar disk.
The infalling core continues to land at the outer edge of the circumstellar disk until the core depletes. The infall rates on the circumstellar disk are in agreement with what can be expected from the free-fall collapse \citep{2010Vorobyov}. Computations continue up to 1.0~Myr, thus covering the entire embedded phase and the early T Tauri phase of disk evolution.

\subsection{Global evolution of gas and dust in the disk}
In this section, we study the formation and long-term evolution of the circumstellar disk.
The top row of Figure~\ref{fig:1} presents the temporal evolution of the azimuthally-averaged gas surface density for models with three values of the $\xi$-parameter: 5\% (left column), 50\% (middle column), and 95\% (right column). Clearly, circumstellar disks in the models with lower $\xi$ are systematically denser than the models with higher $\xi$, particularly in the inner regions within $r \lesssim$~30~au. 
In the $\xi=5\%$ model, the mass transport rate through the CSC onto the star is  only five percent that of the mass transport rate through the sink-disk interface. As a result the matter accumulates in the inner disk until the gas temperature reaches a critical threshold for the thermal ionization and an MRI-triggered accretion bursts sets in. The burst evacuates a larger fraction of the matter accumulated in the CSC and the process repeats itself.  In contrast, in the $\xi=95\%$ model the mass transport rate through the CSC and the mass transport rate at the CSC--disk interface are similar. As a consequence, the matter does not accumulate in the inner disk and the gas surface density is lower in the $\xi=95\%$ model, as is  evident from the isocontours outlining the disk regions with $\Sigma_{\rm g}\ge 10^3$~g~cm$^{-2}$.  We also note that the $\xi=95\%$ model never experiences MRI-triggered accretion bursts, because the gas density and temperature never reach the thermal ionization threshold.
The behavior of small dust particles is similar to that of the gas and is therefore not shown. 

The temporal evolution of grown dust for each model is shown in the second row of Figure~\ref{fig:1}. The black lines present the isocontours with $\Sigma_{\rm d,gr}=10$~g~cm$^{-2}$. The distribution of grown dust is slightly more extended  and the surface density of grown dust is slightly higher in the $\xi=5\%$ model, but the difference in the spatial distribution of grown dust is not as profound as in the case of gas. The similarity between the grown dust distributions is a direct consequence of efficient inward drift of grown dust. The lack of local gas pressure maxima in the disk, as the third row in Figure~\ref{fig:1} demonstrates, allows for grown dust to freely drift through the disk towards the CSC. This also explains a more compact spatial distribution of grown dust as compared to that of gas.

The fourth row of Figure~\ref{fig:1} presents the azimuthally averaged gas temperature in the disk as a function of time. We note that the dust temperature is similar to that of gas in our model. Clearly, the inner disk regions in models with lower $\xi$ are warmer than in models with higher $\xi$ because of higher surface densities and optical depths. 
We also note that the gas temperature in the $\xi=5\%$ model can episodically exceed 1200--1300~K in the inner several AU, triggering accretion bursts as described above (see also Fig.~\ref{arate01}  in Sect.~\ref{accrete}). We note that at these high temperatures dust may sublimate, an effect not taken into account in the current model.

The temporal evolution of the maximal dust grain radius $a_{\rm r}$ is presented in the fifth row of Figure~\ref{fig:1}. Dust grains quickly grow from micron-sized particles  to centimeter-sized pebbles  in the inner several tens of au during just several tens of thousand years after the disk formation instance (which is marked by the horizontal dashed lines). Interestingly, in models with $\xi=50\%$ and $\xi=95\%$ the largest dust grains are concentrated in a ring  at a distance of several tens of au from the star, rather than in the inner disk regions as in the $\xi=5\%$ model. The reason for this distinct behaviour is discussed in more detail further in the text.

Finally, the bottom row of Figure~\ref{fig:1} presents the 
spatial distribution of the total dust-to-gas ratio $\zeta_{\rm d2g}$ vs. time. Significant deviations of $\zeta_{\rm d2g}=(\Sigma_{\rm d, gr}+\Sigma_{\rm d, sm})/\Sigma_{\rm g}$  from the canonical 1:100 value are notable in the inner disk regions. The most striking case is  the $\xi=5\%$ model, which demonstrates a decrease in $\zeta_{\rm d2g}$ by almost an order of magnitude (compared to 1:100) in the inner several au of the disk. Conversely, the other two models show an increase in $\zeta_{\rm d2g}$ by a factor of several near the CSC--disk interface.  

To understand the reason for the decrease of $\zeta_{\rm d2g}$ in the $\xi$=5\% model, we refer to the first and second rows of Figure~\ref{fig:1} showing the surface densities of gas and grown dust, respectively.
Consider first the gas density distribution. Clearly, the $\xi$=5\% model is characterized by  notably higher $\Sigma_{\rm g}$ than the other two models. This difference is caused slow mass transport through the CSC. The resulting ``bottle neck'' leads to the accumulation of large amounts of gas in the disk inner regions. Furthermore, strong negative pressure gradients that develop in the inner disk (see the third row) prevent gas from freely flowing towards the star. Consider now the surface densities of grown dust, which spatial distributions are remarkably similar in models with different $\xi$. It is known that grown dust drifts in the direction of increasing gas pressure. Although small local pressure maxima may be present in the disk in the form of spiral arms, the general trend is evident -- the gas pressure increases inward. As a result, dust freely drifts to the CSC in all three models and small differences in the drift speed may only arise due to variations in the Stokes number. The net effect is that the decrease in $\zeta_{\rm d2g}$ that we witness in the $\xi$=5\% model is {\it not} due to dust depletion, but is due to gas accumulation in the inner disk. The opposite effect is seen in the $\xi$=50\% model and especially in the $\xi$=95\% model -- a fast mass transport through the CSC leads to gas depletion in the inner one au near the CSC--disk interface. As a result, $\zeta_{\rm d2g}$ increases there.

\begin{figure}
\includegraphics[width=\linewidth]{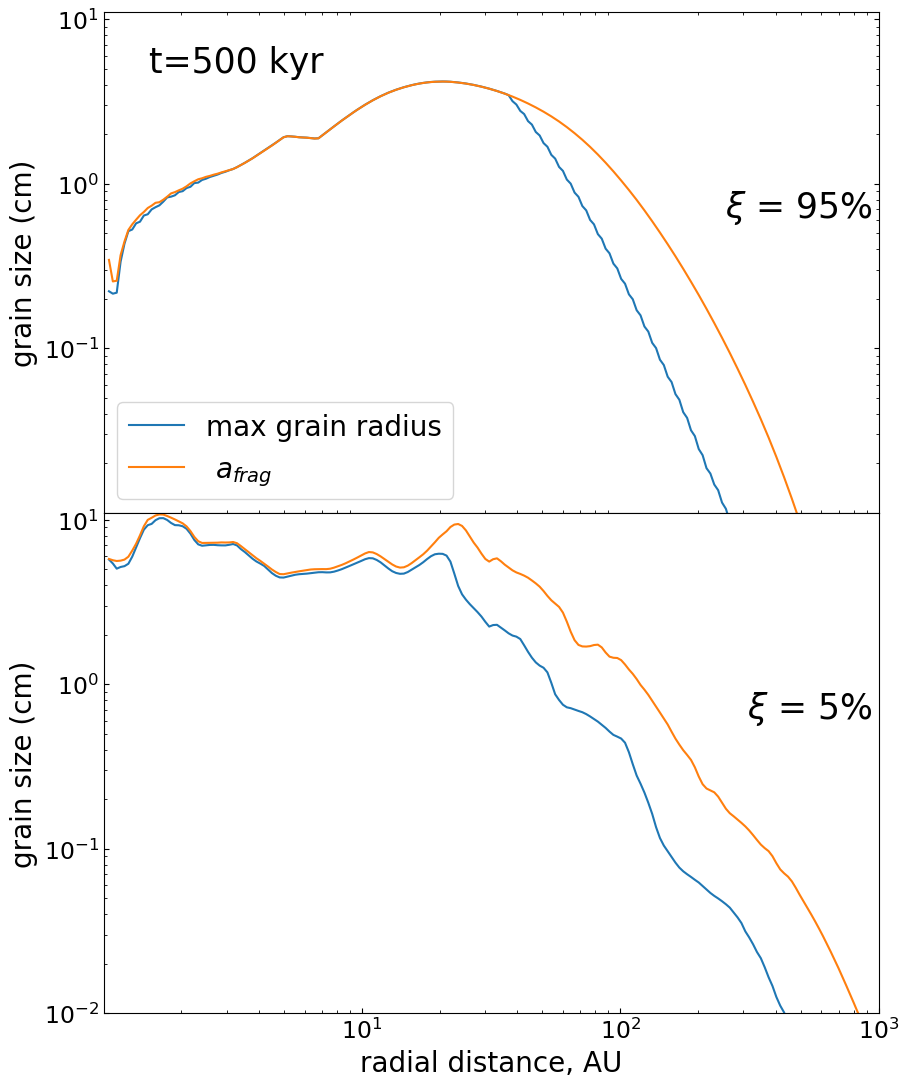}
\caption{Radial profiles of the azimuthally-averaged maximal radius of dust grains and fragmentation limit shown with the blue and orange lines, respectively. The top panel shows the model with $\xi$=95\% and the bottom panel -- the model with $\xi$=5\%. }
\label{fig:2}
\end{figure}

We further analyze the radial distribution of  the maximum grain radius $a_{\rm r}$ in the disk. The blue lines in Figure~\ref{fig:2} present $a_{\rm r}$ vs. radial distance in models with $\xi=95\%$ (top panel) and $\xi=5\%$ (bottom panel). The orange lines show the fragmentation barrier defined as
\begin{equation}
 a_{\rm frag}=\frac{2\Sigma_{\rm g}v_{\rm frag}^2}{3\pi\rho_{\rm s}\alpha c_{\rm s}^2},
 \label{afrag}
\end{equation}
where  $v_{\rm frag}$ is a threshold value for the fragmentation velocity taken to be 30~m~s$^{-1}$ in this study and  $\rho_{\rm s}=2.24$\,g~cm$^{-3}$ the material density of grains. By definition, dust cannot grow in our model beyond the size defined by the fragmentation barrier. The bouncing effects are not considered in our dust growth model. All radial profiles are plotted at $t=500$~kyr.

Two interesting features can be seen in Figure~\ref{fig:2}. First, the dust size is limited by the fragmentation barrier only in the inner several tens of au. In the outer disk regions, the size of dust grains never reaches the fragmentation barrier. This means that the dust growth can be split into two regimes: the fragmentation-limited one in the inner disk and the drift-limited one in the outer disk \citep[e.g.,][]{2016Birnstiel}. Second, the fragmentation barrier in the $\xi=95\%$ model has a well defined peak at 20~au and decreases towards the star, while in the $\xi=5\%$ model the fragmentation barrier has a plateau in the inner 20--30~au. This explains the peculiar spatial distribution of dust sizes in the fifth row of Figure~\ref{fig:1} -- the maximum in $a_{\rm r}$ in the $\xi$=95\% model is reached at several tens of au and not in the innermost disk regions, as might be naively expected.

\begin{figure}
\includegraphics[width=\linewidth]{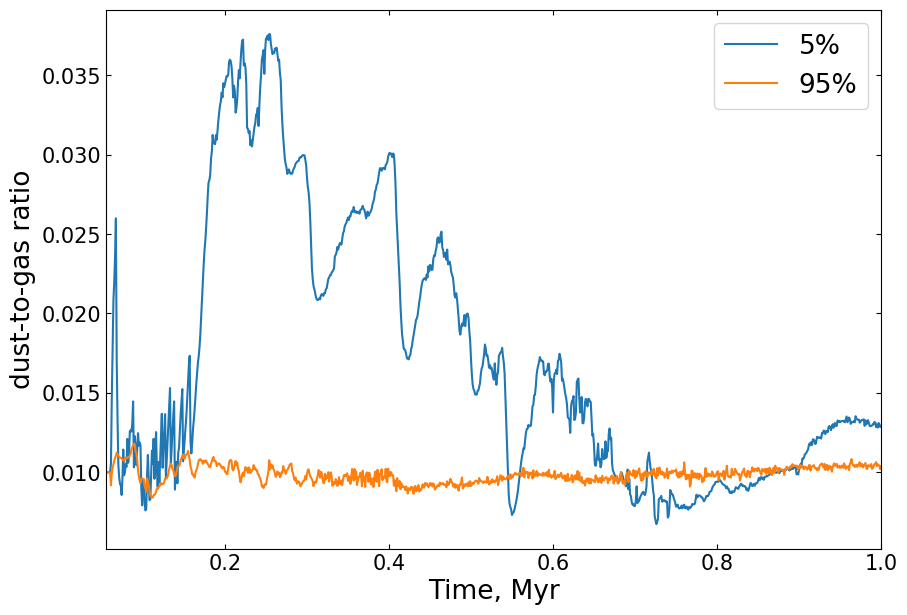}
\caption{Evolution of the dust-to-gas mass ratio in the CSC. The blue and orange lines correspond to the $\xi$=5\% and $\xi$=95\% models, respectively.}
\label{fig:3}
\end{figure}

It is interesting to see how the total dust-to-gas ratio in the CSC evolves with time. Figure~\ref{fig:3} presents the $\zeta_{\rm d2g}$ values in the CSC  as a function of time for the $\xi=95\%$ model (the orange line) and $\xi=5\%$ model (the blue line). In the former model, $\zeta_{\rm d2g}$ stays around 0.01 with small deviations from this canonical value. In the latter model, however, $\zeta_{\rm d2g}$ demonstrates large deviations from the canonical value mainly to higher values. 
In the $\xi=$5\% model we therefore see an interesting trend -- the total dust-to-gas ratio is decreased in the disk region between one and ten~au (see the bottom row in Fig.~\ref{fig:1}), but is enhanced in the CSC (i.e., at $\le$~1.0~au). We note that $\zeta_{\rm d2g}$ may become even higher in the innermost parts of the CSC if additional drift of grown dust takes place there. This, however, cannot be resolved in our numerical model.

\begin{figure}
\includegraphics[width=\linewidth]{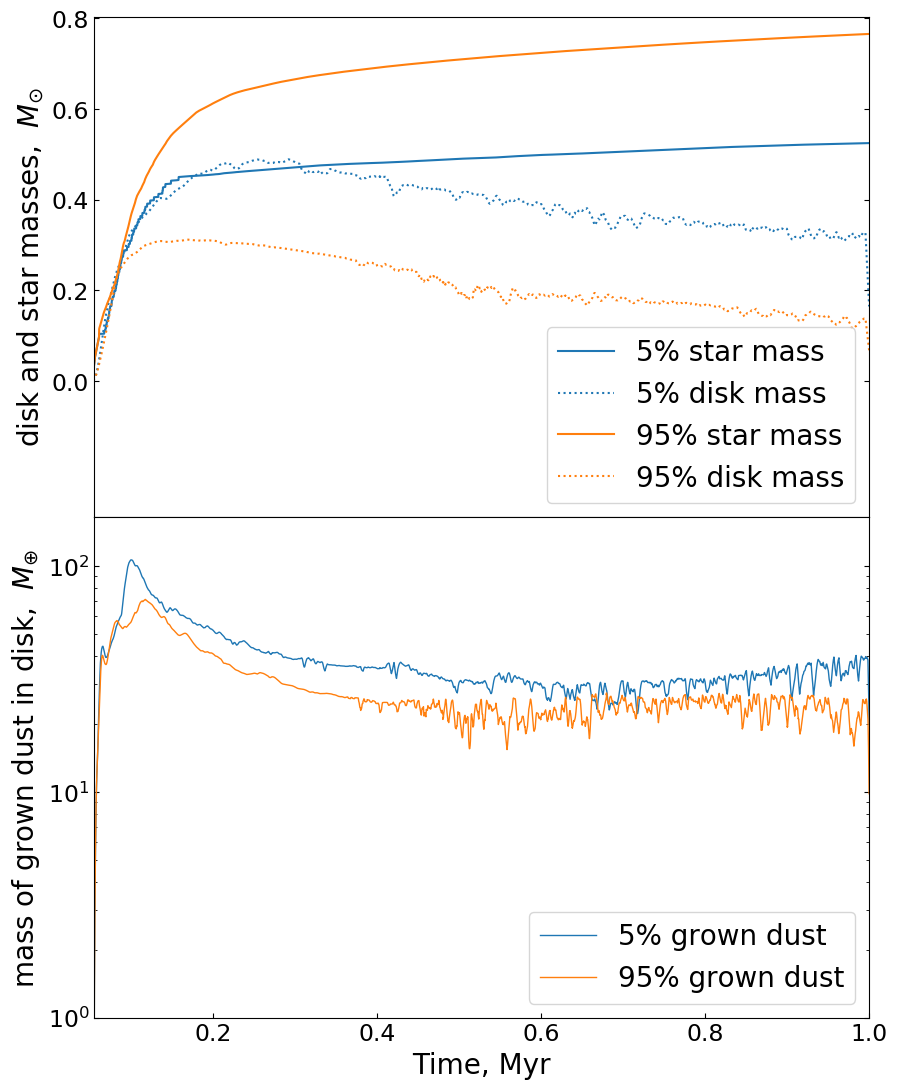}
\caption{Total gas plus dust mass in the disk and stellar mass (top panel) and the mass of grown dust in the disk (bottom panel) in models with $\xi$=5\% (blue lines) and $\xi$=95\% (orange lines). }
\label{fig:7}
\end{figure}

\begin{figure}
\begin{centering}
\includegraphics[width = \linewidth]{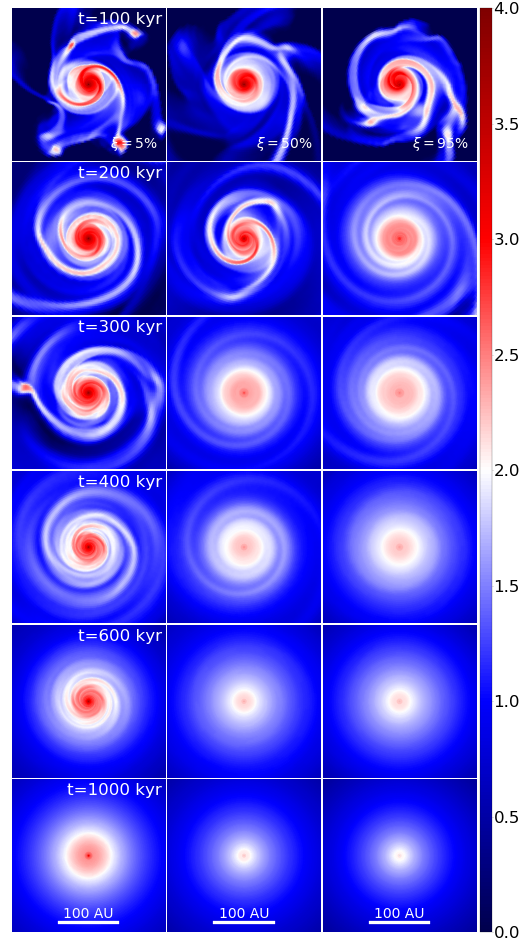}
\par\end{centering}
\caption{Gas surface density maps in the inner $250\times250$~au$^2$ box shown at consecutive evolutionary times for all three models with distinct $\xi$-parameters. Time is indicated in the left column and is counted from the onset of numerical simulations and the disk forms at 56~kyr. The scale bar is in log g~cm$^{-2}$. }
\label{Instab}
\end{figure}

Finally, the models with different rates of mass transport through the CSC are also characterized by distinct disk and stellar masses. In Figure~\ref{fig:7} we present the integrated masses of gas and grown dust in the disk and also the stellar mass as a function of time for the $\xi$=5\% model (blue lines) and $\xi$=95\% model (orange lines). Clearly, the $\xi=95\%$ model has a systematically higher stellar mass, but lower gas disk mass than its $\xi=5\%$ counterpart. Higher stellar masses and lower disk masses make gravitational instability less efficient (see Sect.~\ref{gravinst} and Fig.~\ref{Instab}).
The mass of grown dust in the disk, however, differs only by a factor of unity between the models, which is a consequence of fast and unhindered inward radial drift of grown dust in both models.
We also note that the total mass of grown dust in both models hardly reaches several tens of Earth masses beyond 1.0 au, a rather low value for efficient planet formation to set in -- a feature also noted by \citet{2018VorobyovAkimkin}.

\subsection{Protostellar accretion}
\label{accrete}
Mass accretion on the nascent protostar plays an important role in the dynamical and chemical evolution of a young stellar object. Accretion brings matter, angular momentum, and energy to the star, thus affecting its internal structure. Accretion also feeds back to the disk via accretion luminosity, providing the main source of disk heating in its intermediate and outer regions (including the envelope in the embedded phase). Determining the character of protostellar accretion is therefore one of the major problems  in the theory of  star formation.

\begin{figure}
\includegraphics[width=\linewidth]{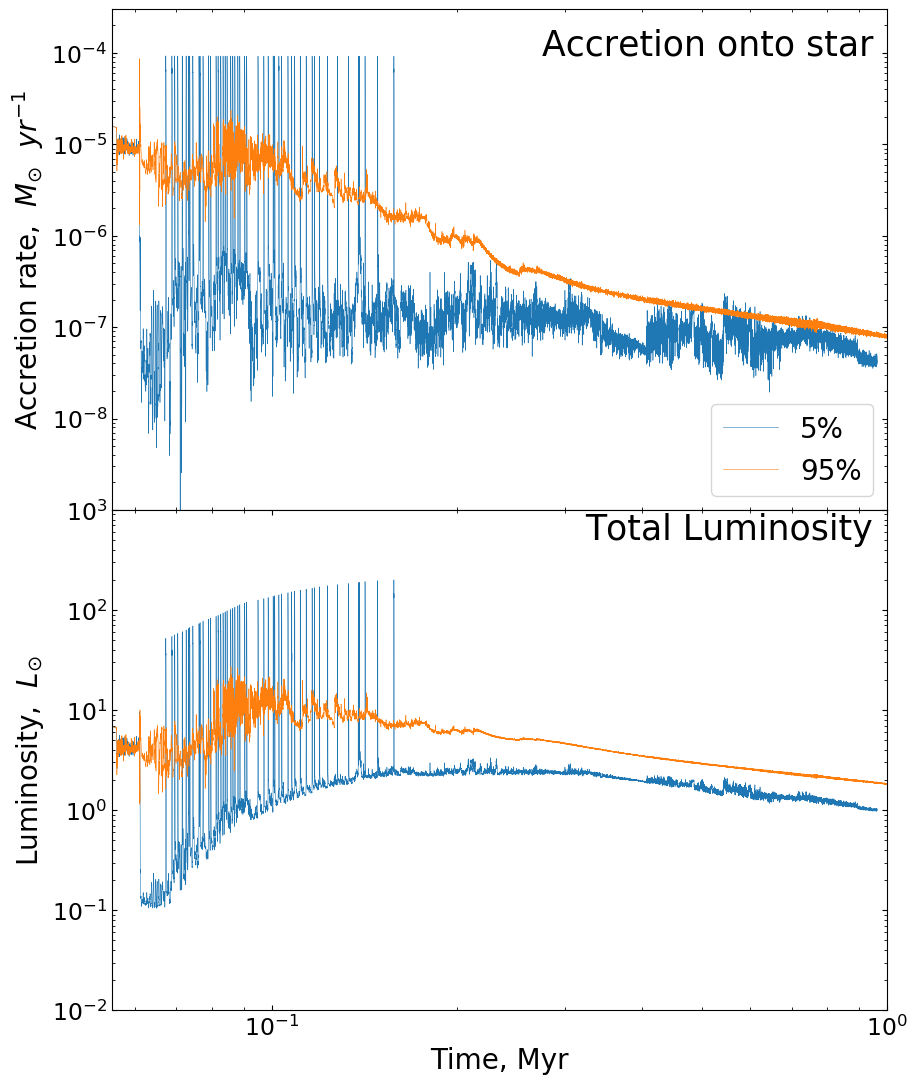}
\caption{Accretion rate (top panel) and total stellar luminosity (bottom panel) for the $\xi=5\%$ model (blue lines) and $\xi=95\%$ model (orange lines).}
\label{arate01}
\end{figure}

\begin{figure*}
\includegraphics[width=\textwidth]{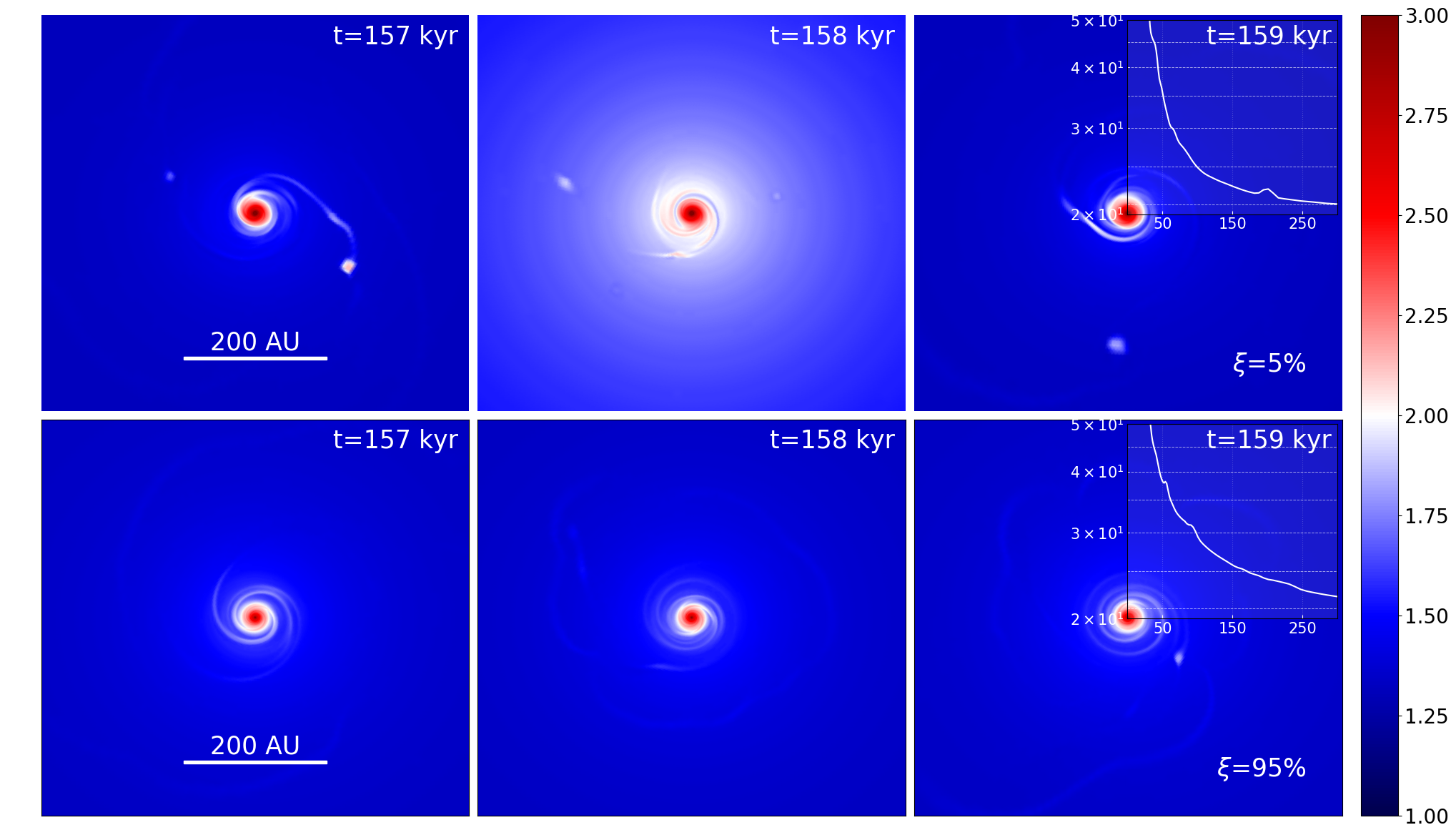}
\caption{\textbf{Top row.} Spatial maps of the gas temperature in the inner $600\times 600$ au$^{2}$ box before an accretion burst (left panel), during the burst (middle panel), and after the burst (right panel) for the $\xi=5\%$ model. \textbf{Bottom row.} Spatial maps of the gas temperature in the $\xi=95\%$ model characterized by steady accretion without bursts. Same evolutionary times as for the $\xi=5\%$ model are shown. The inset plots in the right column present the azimuthally-averaged temperature in the same spatial scale as the 2D temperature maps. The scale bar is in log Kelvin.}
\label{fig:6}
\end{figure*}

Figure~\ref{arate01} shows the mass accretion rates on the star $\dot{M}_\ast = \dot{M}_{\ast,\rm csc} + \dot{M}_{\ast,\rm bst}$ and the total stellar luminosity $L_{\rm tot}$ (accretion and photospheric) in models with $\xi=5\%$ and $\xi=95\%$. 
Clearly, $\dot{M}_\ast$ in the $\xi=5\%$ model is highly variable in the early 0.15~Myr of evolution and shows frequent accretion bursts  alternated with longer periods of quiescent low-rate accretion. At the same time, the $\xi=95\%$ model is characterized by a gradually declining accretion rate with only low-amplitude flickering. Moreover, $\dot{M}_\ast$ in the $\xi=95\%$ model is notably higher than the quiescent accretion rate in the $\xi=5\%$ model (the latter stays around $10^{-7}~M_\odot$~yr$^{-1}$). A similar trend stands for the total stellar luminosity -- $L_{\rm tot}$ in the $\xi$=5\% model features strong outbursts, while the $\xi$=95\% model has only low-amplitude variations. The latter model, nevertheless, is characterized by a higher $L_{\rm tot}$ in the quiescent mode.

This difference in the character of mass accretion and stellar radiative output may have important consequences for the stellar and disk evolution. Accretion bursts affect the physical characteristics and chemical composition of young protostars, causing peculiar excursions on the Hertzsprung-Russel diagram and changing their lithium abundance \citep{2019ElbakyanVorobyov,2017BaraffeElbakyan}. In addition, luminosity bursts raise the disk temperature, thus affecting its dynamics and chemical composition \citep{2011Stamatellos,2018MolyarovaAkimkin}. Finally, we note the accretion bursts in the $\xi=5\%$ model, though being quite intense and numerous, do not make up for the quiescent accretion and the final stellar mass in this model is notably lower than that of the $\xi=95\%$ model without the bursts.

Figure~\ref{fig:6} presents the gas temperature distribution in the $600 \times 600$~au$^2$ box in models with $\xi=5\%$ (top panels) and $\xi=95\%$ (bottom panels). Three time instances are chosen to show the temperature distribution in the $\xi$=5\% model  before the burst ($t=157$~kyr), during the burst ($t=158$~kyr), and after the burst ($t=159$~kyr). The $\xi$=95\% model has no bursts, but its temperature distribution is shown at the same times for comparison. We note that the isolated blobs of hot gas that are seen in the upper panels are gaseous clumps formed via disk gravitational fragmentation. Clearly, during the burst the gas temperature exceeds the pre- and post-burst values by a factor of several. 
Furthermore, the insets in the right column  demonstrate that the disk is warmer in the $\xi=95\%$ model and is colder in the $\xi=5\%$ model during the quiescent phase of accretion. This difference in gas temperatures has an important consequence for the longevity of disk gravitational instability, as we will see later in Sect.~\ref{gravinst}.
We conclude that the rate of mass transfer through the CSC has a significant effect on the character of mass accretion and luminosity of the nascent protostar.

\subsection{Gravitational instability and fragmentation}
\label{gravinst}
Gravitational instability and fragmentation are among the most important physical phenomena that occur in the early evolution of protostellar disks. Gravitational instability is a major mass and angular momentum transport mechanism in the embedded phase of disk evolution \citep{VorobyovBasu2009}, while gravitational fragmentation is considered as a likely gateway for the formation of giant planets and brown dwarfs \citep[][]{2003Boss, 2007MayerLufkin, 2009StamatellosWhitworth, 2011ForganRice, 2013Vorobyov, 2015Meru, 2016Vorobyov,2017Nayakshin}. In addition, disk fragmentation can trigger accretion bursts caused by inward migration and infall of gaseous clumps on the star \citep[e.g.,][]{2005VorobyovBasu,2014Machida,VorobyovBasu2015,2017MeyerVorobyov,2018ZhaoCaselli,2018VorobyovElbakyan}. Gravitational instability and fragmentation occurs in the intermediate and outer disk regions and it appears at a first glance that physical processes in the CSC are unlikely to influence this phenomenon. However, as we saw in Figs.~\ref{fig:1} and \ref{fig:7}, the rate of mass transport through the CSC can affect the disk mass, gas density and temperature distributions in the bulk of the disk, which in turn can have implications for the development of gravitational instability and fragmentation.

\begin{figure}
\includegraphics[width = \linewidth]{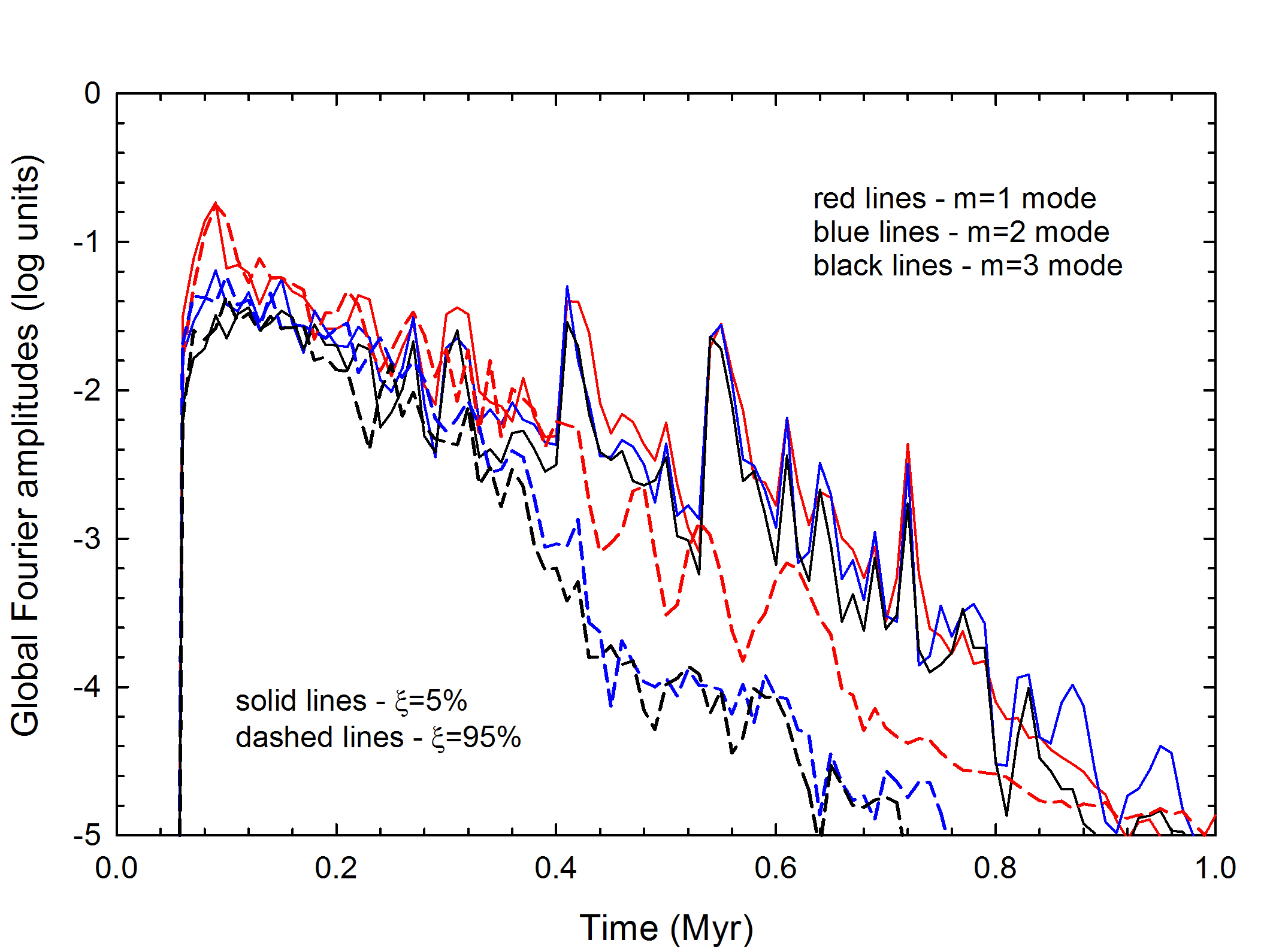}
\caption{Global Fourier amplitudes. The solid and dashed lines of distinct colour correspond to models with slow ($\xi=5\%$) and fast ($\xi=95\%$) mass transport through the CSC, respectively. Fourier amplitudes for the first three modes ($m=1$, $m=2$, and $m=3$) are shown.}
\label{fourier}
\end{figure}

\begin{figure}
\includegraphics[width=\linewidth]{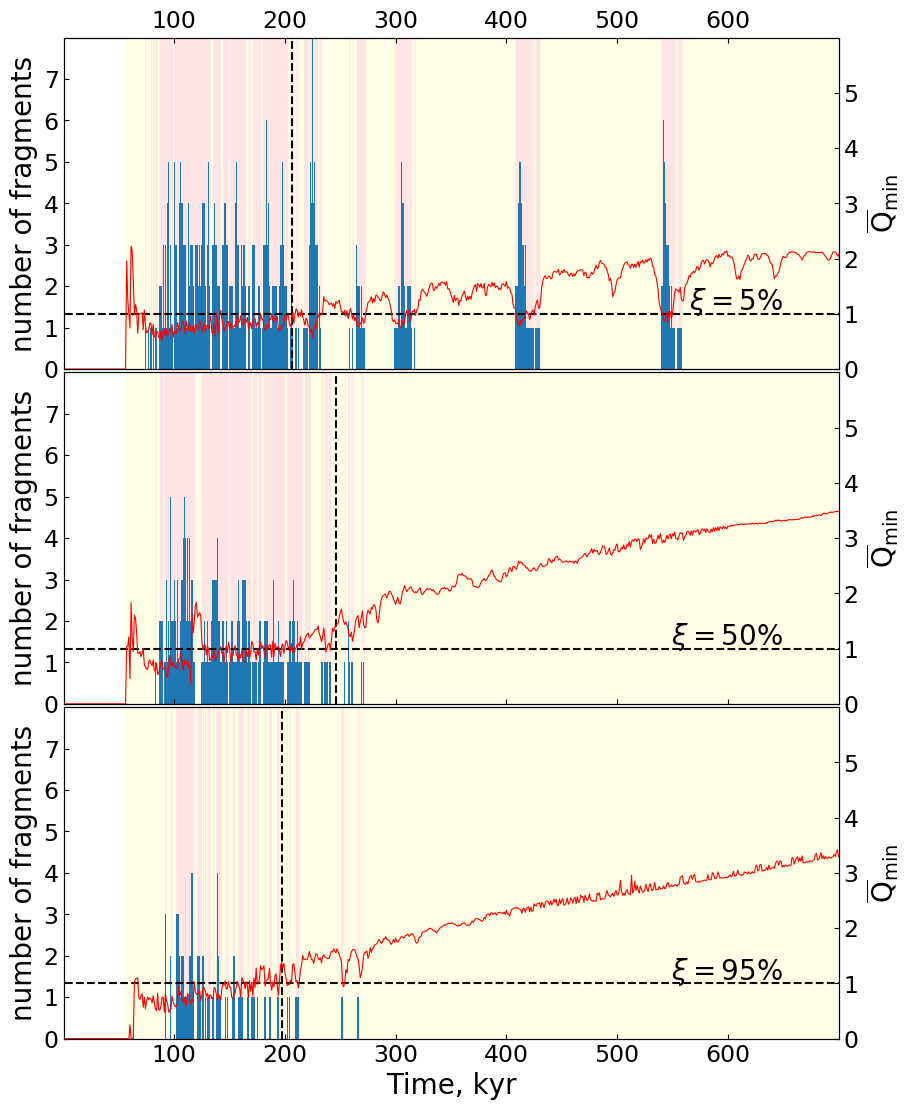}
\caption{Total number of fragments present in the entire disk as a function of time in the $\xi$ = 5\% model (top panel), $\xi$=50\% model (middle panel), and $\xi=95\%$ model (bottom panel). The vertical dashed lines mark the onset of the Class II phase of disk evolution. Pink zones correspond to time periods in the Class II phase when fragments exist in the disk, and yellow zones show the time periods without any fragments in the disk. The red lines show $\overline Q_{\rm min}$ -- the mean minimal Toomre parameter in the disk.}
\label{clnum}
\end{figure}

Figure~\ref{Instab} presents the gas surface density distributions in the inner $250\times250$~au$^{2}$ box in all three considered models with distinct $\xi$-parameters at consecutive evolutionary times as indicated in the first column. More specifically, the first, second, and third columns present the models with $\xi$=5\%, $\xi$=50\%, and $\xi$=95\%, respectively. The early evolution in all models is qualitatively similar -- the disk is characterized by strong gravitational instability. Fragmentation is also evident in several panels. However, the subsequent evolution demonstrates notable differences. For instance, the $\xi$=95\% model has a nearly axisymmetric disk already at 400~kyr, while the other models have a well-defined spiral structure at this evolutionary time.  This difference in the longevity of gravitational instability can in part  be explained by the apparent difference in the disk and stellar masses in these models. As Fig.~\ref{fig:7} demonstrated, the disk mass is the lowest and the stellar mass is the highest in the $\xi$=95\% model, both acting to stabilize the disk against gravitational instability. In addition, a systematically higher gas temperature of the disk in the $\xi$=95\% model works for faster disk stabilization.

The strength of gravitational instability can be quantified using the Fourier amplitudes defined as 
\begin{equation}
C_{m} (t) = {1 \over M_{\rm d}} \left| \int_0^{2 \pi} 
\int_{r_{\rm sc}}^{R_{\rm d}} 
\Sigma_{\rm g}(r,\phi,t) \, e^{im\phi} r \, dr\,  d\phi \right|,
\label{fourierEq}
\end{equation}
where $M_{\rm d}$ is the disk mass, $R_{\rm d}$ is the diskâs physical outer radius, and $m$ is the azimuthal wave number of the disk's spiral mode. 
When the disk surface density is axisymmetric, the amplitudes of all modes are equal to zero. When, for example, $C_m (t)$ = 0.1, the perturbation amplitude of spiral density waves in the disk is 10\% that of the underlying axisymmetric density distribution.

Figure~\ref{fourier} presents the global Fourier amplitudes as a function of time for the $m=1$, $m=2$, and $m=3$ spiral modes. The solid and dashed lines correspond to the $\xi$=5\% 
and $\xi$=95\% models, respectively. Both the models are characterized by $C_m$ that are highest in the early evolution but decrease with time, reflecting a gradual decline in the strength of gravitational instability. The decline is however the fastest in the $\xi$=95\% model, in agreement with the fastest stabilization of the disk discussed in the context of Figure~\ref{Instab}. We also note an interesting feature seen in the $\xi$=5\% model -- episodic surges in $C_m$ are apparent between 0.3~Myr and 0.7~Myr. The nature of these events is discussed in more detail below.

\begin{figure*}
\includegraphics[width=\textwidth]{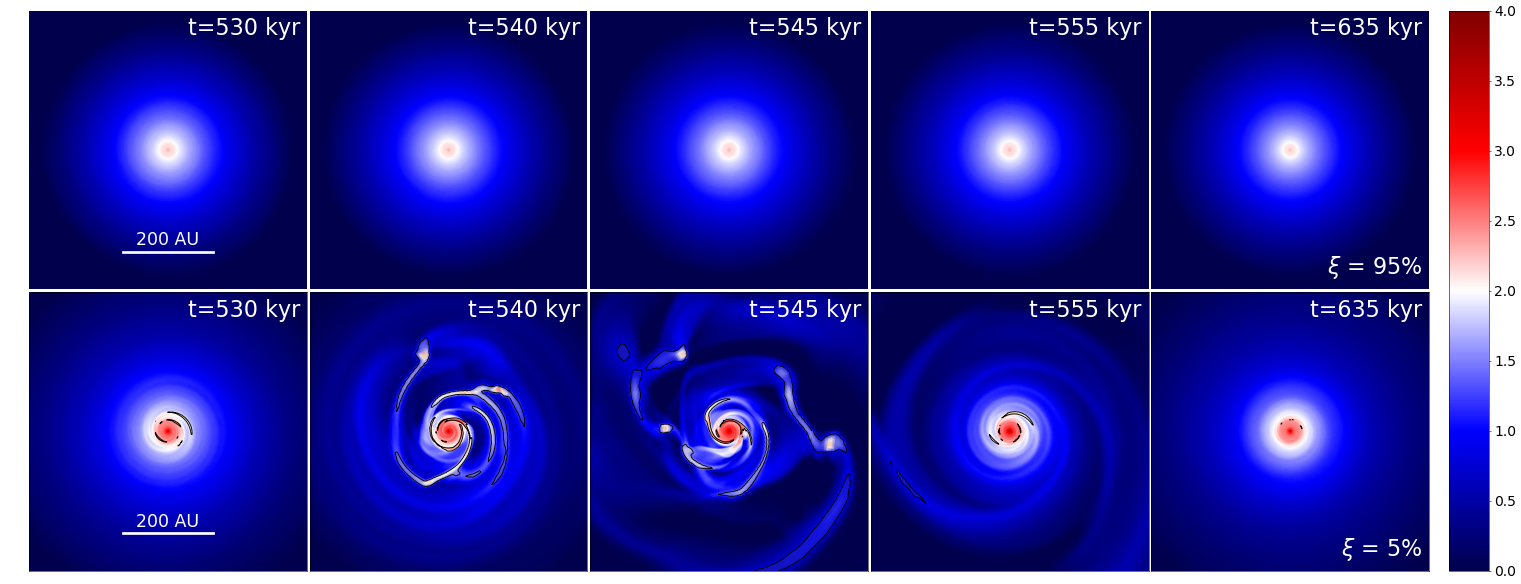}
\caption{Gas surface density maps in the inner $600\times600$~au$^2$ box shown during a short time period between 530~kyr and 635~kyr for the $\xi=95\%$ model (top panels) and $\xi=5\%$ model (bottom panels). The black contour lines outline the disk regions where the Toomre Q-parameter is smaller than unity. The time is counted from the onset of numerical simulations. The disk is formed at 56~kyr. The scale bar is in log g cm$^{-2}$.}
\label{fig:4}
\end{figure*}

We have already pointed out that our models exhibit not only gravitational instability, but also gravitational fragmentation.
In Figure~\ref{clnum} we plot the total number  of fragments that simultaneously exist in the disk for the $\xi$ = 5\% (top panel), $\xi$=50\% model (middle panel), and $\xi=95\%$ models (bottom panel).  
We distinguish the newly formed fragments using the fragment-tracking method described in detail in \citet{2018VorobyovElbakyan}. This method requires that the fragment is pressure-supported, having a negative pressure gradient with respect to the center of the
fragment, and also that the fragment is kept together by gravity, with 
the potential well being deepest at the center of the fragment. 
We first scan the disk for the local density enhancements above a certain density limit. If any of the two conditions fails 
at the center of the density enhancement, then this candidate is rejected,
meaning that we falsely took a local density perturbation for a pressure-supported, gravity-bound fragment. If both conditions are fulfilled at the center of the density enhancement, we continue marching 
from its center and check the neighbouring cells until any of 
the two conditions is violated. The grid cells that fulfill both conditions constitute the fragment.

Let us first consider the $\xi$=5\% model. The first fragment forms as soon as 80~kyr after the onset of the simulation (or 25~kyr after disk formation) and is followed by an uninterrupted series of fragmentation events lasting for $\approx 150$~kyr, during which fragments are constantly present in the disk. The maximal number of fragments existing simultaneously in the disk is eight. In the following evolution, the disk enters an episodic fragmentation mode with major fragmentation episodes occurring at 270, 300, 410, and 530~kyr. In contrast, disk fragmentation in the $\xi$=50\% and $\xi=95\%$  models is less pronounced. It starts later, ends sooner, and forms less fragments at a time than in the $\xi=5\%$ model.  The episodic fragmentation mode is suppressed -- fragmentation ends after 270~kyr and no fragments are found in the later evolution. This difference in the intensity of disk fragmentation is concordant with the corresponding difference in the strength of gravitational instability (see Fig.~\ref{Instab}) and is caused by distinct disk masses, stellar masses, and disk temperatures in these three models, as was discussed above.

Interestingly, the episodic fragmentation events take place in the Class II phase of disk evolution, the onset of which is marked by the dashed vertical line in Fig.~\ref{clnum}) and is defined as the time instance when less than 5\% of the initial mass reservoir of the pre-stellar core is left in the infalling envelope. It is usually thought that disk fragmentation is most likely to develop in the embedded phase of disk evolution when mass-loading from the envelope replenishes the disk mass loss via accretion and sustains the disk in the unstable mode \citep[e.g.][]{2005VorobyovBasu,2008Kratter}.
Our numerical simulations indicate that disk fragmentation is also possible in more evolved disks when mass-loading is no more operational.
We define the evolutionary state when at least one fragment exists  in the disk as the ``fragmented-disk state''. Conversely, the ``stable-disk state'' denotes the  state when no fragments are found in the disk. We note that the lifetime of the fragments may be underestimated because of limited numerical resolution, which may cause their premature destruction via tidal torques. Thus, the duration of the fragmented-disk state is to be considered as a lower limit.

The fragmented-disk and stable-disk states are highlighted in Fig.~\ref{clnum} with the pink/yellow background color, respectively.
Clearly, the time spent in the stable-disk state is longer than in the fragmented-disk state. When considering only the early T Tauri phase (200--700~kyr), the $\xi=5\%$ model spends 80\% of in the stable-disk state and only 20\% in the fragmented-disk state.
This means that detecting a Class II disk in the stable-disk state is more likely than in the fragmented-disk one, especially considering that disk fragmentation ends after about 0.5~Myr of disk evolution.

A useful quantity that describes the disk propensity to gravitational instability and fragmentation is the Toomre $Q$-parameter. For a mixture of gas and dust the $Q$-parameter can be defined as
\begin{equation}
Q={\tilde{c}_{\rm s} \Omega \over \pi G (\Sigma_{\rm g}+\Sigma_{\rm d,sm} + \Sigma_{\rm d,gr})},
\label{ToomreQ}
\end{equation}
where  $\tilde{c}_{\rm s}=c_{\rm s}/\sqrt{1+\zeta_{\rm d2g}}$ the modified sound speed 
in the presence of dust. 
As follows from its definition, the Toomre parameter depends on the local characteristics in the disk and can vary throughout the disk extent. 
To relate the fragmented-disk state with the $Q$-parameter, we need to choose a representative $Q$-value for the entire disk. Since the disk propensity to gravitational instability is inversely proportional to the $Q$-value, we search for the minimal values of $Q$ at every radial annulus in the disk and then arithmetically average the resulting minimal values to derive the mean minimal Toomre parameter $\overline Q_{\rm min}$ for the disk. We exclude the inner 15~au of the disk from this procedure, since this region is characterized by $Q$-values that are much higher than 2.0 due to high temperature and strong shear. The resulting quantity is plotted in Fig.~\ref{clnum} with the red lines. In the fragmented-disk state the $\overline Q_{\rm min}$-values drop often below 1.0, while in the stable-disk mode they rise above 2.0, which is consistent with the theoretical expectations.

To further investigate the episodic mode of disk fragmentation,  we plot in Figure~\ref{fig:4} the gas surface density distributions in the $\xi=95\%$ model (top row) and $\xi=5\%$ model (bottom row) during an evolutionary period of 105~kyr starting from $t=530$~kyr
after the onset of pre-stellar core collapse. At this evolutionary stage, the disk is about 480~kyr old and is in the T Tauri stage of evolution, where gravitational fragmentation is not expected to be a dominant mode of evolution. Indeed, in the $\xi=95\%$ model the disk is smooth, axisymmetric, and gravitationally stable. At the same time,  the disk in the $\xi=5\%$ model experiences a transitory episode of vigorous  gravitational instability and fragmentation.  The black contour lines delineate disk regions where the Toomre $Q$-parameter 
is smaller than unity. 
Clearly, the Q-parameter drops below unity only in the $\xi=5\%$ model showing disk fragmentation, in agreement with the theoretical predictions. 

\begin{figure}
\includegraphics[width=\linewidth]{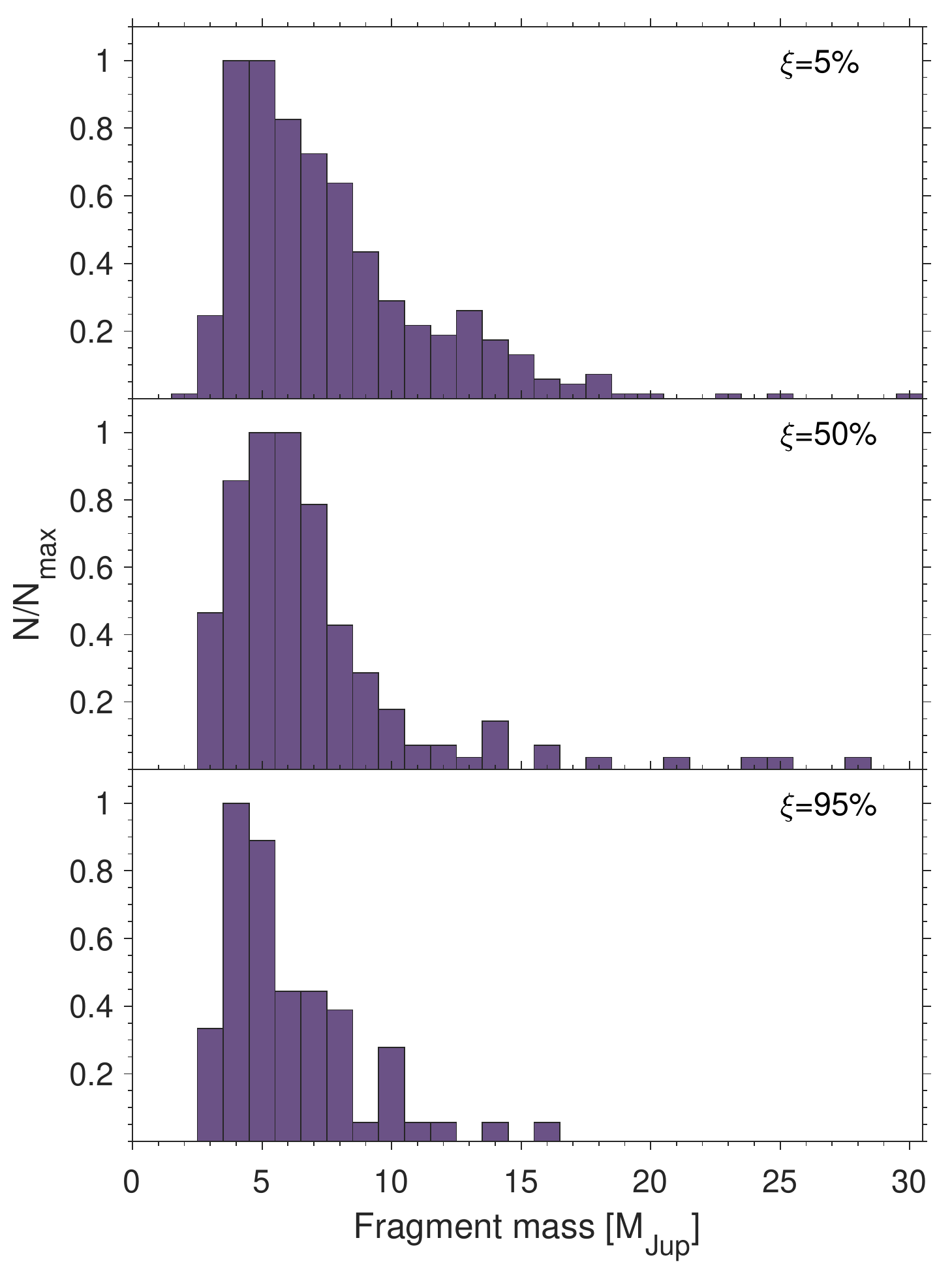}
\caption{ Normalized number of fragments in the disk vs. fragment mass in the $\xi=5\%$ model (top panel), $\xi=50\%$~model (middle panel), and $\xi=95\%$~model (bottom panel). The normalization is done to the peak value of the distribution function of fragment masses.}
\label{num_frag}
\end{figure}

The diagrams in  Figure~\ref{num_frag} presents the normalized number of fragments as a function of their mass in the $\xi=5\%$ model (top panel), $\xi=50\%$ model (middle panel), and $\xi=95\%$ model (bottom panel). We searched for the presence of fragments in the disk every 1.0~kyr of disk evolution, totaling to almost 1000 time snapshots. We then used these snapshots to calculate the total number of fragments that were present in the disk during its entire evolution. Since the lifetime of a fragment in the disk may be longer than 1.0~kyr, this procedure implies that some fragments may have been counted several times. Because of this reason we provided the normalized number of fragments and not their absolute number. This procedure also means that our diagrams favour fragments with longest lifetimes, which, in our opinion, is sensible. Clearly, the peak of the mass distribution is at 4--6~$M_{\rm Jup}$ and the number of fragments drops fast on both sides of the peak. The lowest mass is 2~$M_{\rm Jup}$ and the highest is 30~$M_{\rm Jup}$. If we set an upper mass limit of 13~$M_{\rm Jup}$ for a giant planet, then most fragments are in the planetary mass regime.

The numerical resolution of our simulations does not allow us to reliably follow  the fate of the fragments as they are destroyed by tidal torques. However, higher resolution simulations demonstrated that fragments may survive and form massive giant planets and brown dwarfs on wide orbits \citep[e.g..][]{2018VorobyovElbakyan}.
It remains to be understood what consequences the delayed formation of giant planets and/or brown dwarfs may have on the structuring of planetary systems. We conclude that disk fragmentation is a likely phenomenon in the evolution of massive circumstellar disks, but its intensity and longevity depend sensitively on the value of $\xi$. We plan to address the episodic fragmentation mode in more detail in follow-up higher-resolution studies.

\begin{figure}
\includegraphics[width=\linewidth]{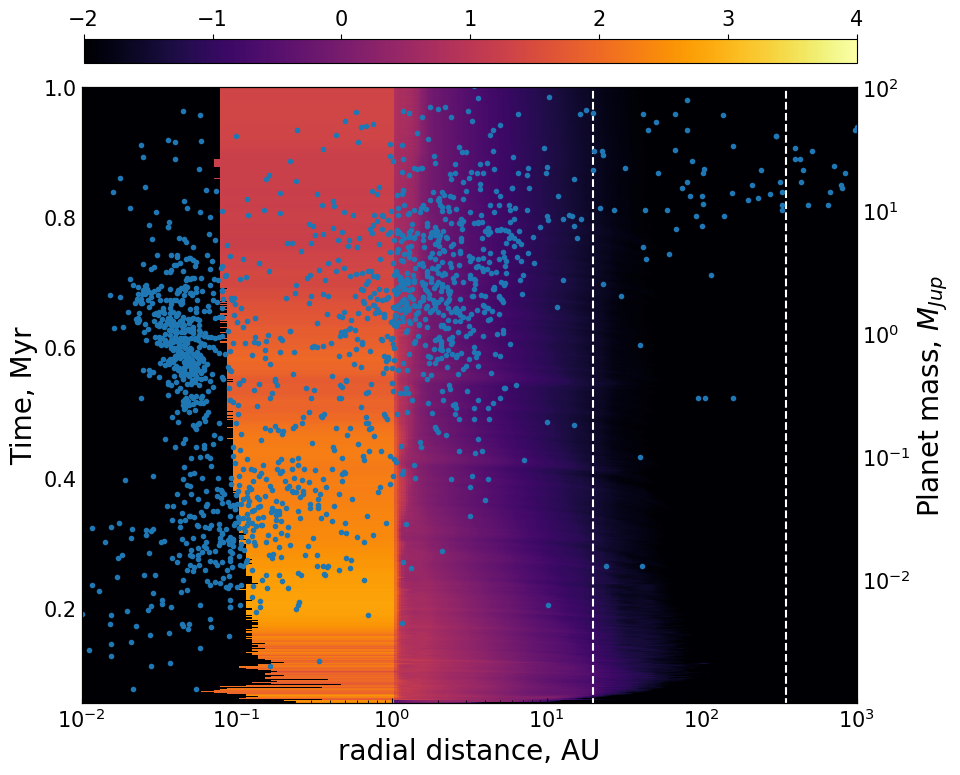}
\caption{Comparison of the orbital positions of known exoplanets with the spatial distribution of grown dust as predicted by our numerical hydrodynamics models. The azimuthally averaged surface density of grown dust is plotted as a function of time using a coloured map. The scale bar on the top shows $\Sigma_{\rm d, gr}$ in log g~cm$^{-2}$. The orbital positions of known exoplanets (taken from http:\textbackslash\textbackslash exoplanet.eu) are indicated with the blue circles as a function of their mass. The vertical dotted lines delineate the disk region where gas giants can be formed via gravitational fragmentation.}
\label{exoplanets}
\end{figure}

\subsection{Orbital diversity of exoplanets}
The known exoplanets demonstrate a surprising diversity in their characteristics, such as mass, density, atmospheric composition, etc. In particular, the orbital positions of exoplanets span a range from $\la 0.1$~au to several hundreds of au. In this section, we compare our model predictions with the orbital distribution of known exoplanets.

The coloured map in Figure~\ref{exoplanets} shows the azimuthally averaged distribution of grown dust vs. time.  We took an arithmetic averaging of $\Sigma_{\rm d, gr}$ in models with three different values of $\xi$. The surface density of grown dust in the CSC is calculated as a ratio of the known mass of grown dust in the CSC (see Eq.~\ref{csc1}) to the surface area of the CSC. The outer boundary of the CSC is 1.0 au and the inner boundary of the CSC is set equal to the dust sublimation radius $r^{\rm in}_{\rm csc}$. The latter is calculated as:
\begin{equation}
r^{\rm in}_{\rm csc} = \left(\frac {L_{\rm tot}}{16 \pi \sigma T_{\rm sub}^4} \frac{\kappa_{P}^{\rm star}}{\kappa_{P}^{\rm sub}}\right)^{1/2},
\label{eq_rsub}
\end{equation}
where $L_{\rm tot}$ is the total stellar luminosity provided by our models, $\sigma$ is the Stefan-Boltzmann constant, $T_{\rm sub}$ is the dust sublimation temperature assumed to be equal to 1500~K, $\kappa_{P}^{\rm star}$ is the Planck mean opacity calculated for the effective temperature of the central star, which in turn is determined from the stellar evolution tracks, and $\kappa_{P}^{\rm sub}$ is the Planck mean opacity calculated for the dust sublimation temperature. We note that $\kappa_{P}^{\rm sub}$ is constant, but $\kappa_{P}^{\rm star}$ changes with time as the central star evolves.   We also note that Eq.~(\ref{eq_rsub}) does not consider that dust grains can be back-heated by dust at larger radii, in which case $r^{\rm in}_{\rm csc}$ may increase by a factor of order unity. 
Two vertical dotted lines in Figure~\ref{exoplanets} highlight the disk regions within which the formation of giant planets via disk gravitational fragmentation is feasible \citep[e.g.,][]{2013Vorobyov,2015Stamatellos,2018VorobyovElbakyan}. The orbital positions of exoplanets as a function of their mass are indicated with the blue circles. We note that a jump in the dust surface density at $r\le1.0$~au is caused by inward drift and accumulation of grown 
dust in the CSC.

Let us first consider the outer parts of the disk. Figures~\ref{Instab} and \ref{fig:4} demonstrate that gravitational instability and fragmentation are operational at distances from tens to hundreds of au. Because of insufficient numerical resolution of our models, the forming fragments are destroyed by tidal torques when migrating towards the star. However, higher resolution simulations indicate that a small fraction of the fragments may survive, open a gap, and settle on orbits from tens of au to hundreds of au, thus giving birth to giant planets on wide orbits \citep[e.g.,][]{2013Vorobyov,2015Stamatellos,2018VorobyovElbakyan}. The range of planetary orbits that may be accounted for by disk gravitational fragmentation is marked by two dashed vertical lines in Figure~\ref{exoplanets}. Moreover, the role of gravitational fragmentation may not be limited to the formation of giant planets on wide orbits. Tidal downsizing of gaseous envelopes of giant planets may turn them into icy giants and even in super-Earths, and these lighter planetary embryos may continue migrating towards the star, creating a population of close-orbit planets \citep[see a review by][]{2017Nayakshin}.

\begin{figure}
\includegraphics[width=\linewidth]{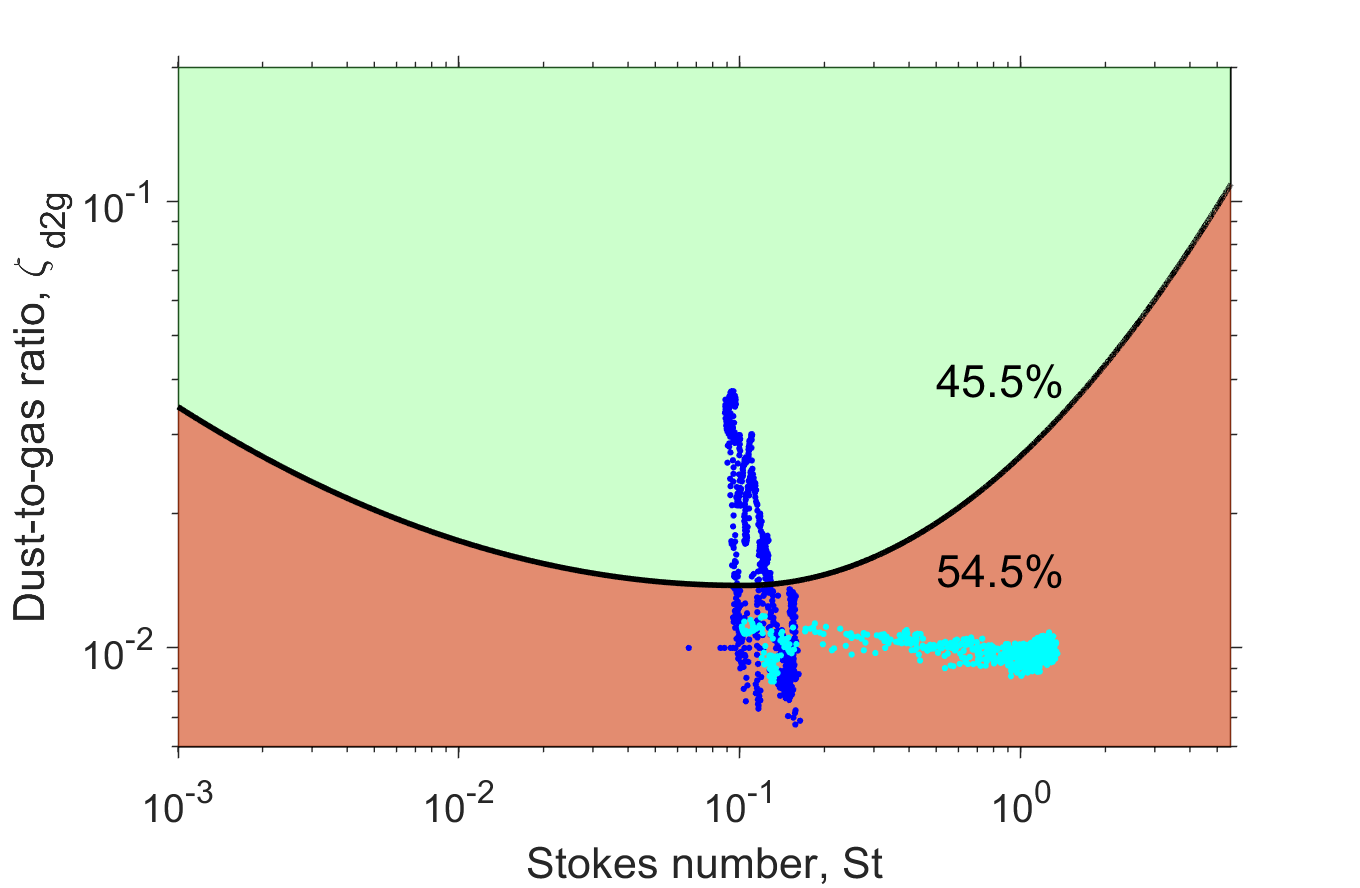}
\caption{Regions in the $\zeta_{\rm d2g}$--St phase space that are subject to the development of streaming instability (highlighted in the green color) and that are stable against streaming instability (highlighted in the brown color). The corresponding data are taken from \citet{2017Yang}. The blue dots present our data for the $\xi=5\%$ model and the numbers show the percentage of time spent by this model in the unstable and stable regimes. The cyan dots present the corresponding data for the $\xi=95\%$ model.}
\label{stream_cond}
\end{figure}

Let us now consider the innermost parts of the disk where the highest accumulation of grown dust is found.  In our models, dust does not grow beyond the fragmentation barrier $a_{\rm frag}$ (see Eq.~\ref{afrag}), effectively limiting the dust size to centimeters or decimeters at most. However, the streaming instability is often considered as a mechanism that helps to overcome the fragmentation barrier, thus assisting the formation of planetesimals and planetary cores in protoplanetary disks \citep[e.g.,][]{2007Johansen}. We do not model the gas dynamics in the CSC, but it is possible to check if conditions for the development of streaming instability are satisfied in the CSC. For these conditions, we choose those suggested by the study of \citet{2017Yang}. The black solid line in Fig.~\ref{stream_cond} divides the $\zeta_{\rm d2g}$--St phase space into the region prone to streaming instability  (highlighted in the green color) and the region where streaming instability is suppressed (highlighted in the brown color). 
Here, $\zeta_{\rm d2g}$ is the ratio of total dust to gas surface densities and St is the Stokes number defined as 
\begin{equation}
 {\rm St}=\frac{\Omega_{\rm K}\rho_{\rm s} a_{\rm r}}{\rho_{\rm g}c_{\rm s}},
 \label{StokesN}
\end{equation}
where $\rho_{\rm g}$ is the gas volume density calculated as $\rho_{\rm g}=\Sigma_{\rm g}/(\sqrt{2\pi} {H_{\rm g}})$, $\rho_{\rm s}=2.24$\,g~cm$^{-3}$ the material density of grains, and $\Omega_{\rm K}$ is the Keplerian velocity. The maximum radius of dust grains $a_{\rm r}$ in the CSC is assumed to be equal to the fragmentation radius $a_{\rm frag}$, which is a legitimate assumption for the inner fragmentation-limited disk regions (see Fig.~\ref{fig:2}). For the speed of sound $c_{\rm s}$ we take the corresponding values at the CSC-disk interface. We note that $\zeta_{\rm d2g}$ can be directly derived from our model of the CSC (see Sect.~\ref{sinkmodel}).

The blue circles in Fig.~\ref{stream_cond} present $\zeta_{\rm d2g}$ vs. St calculated for the CSC in the $\xi=5\%$~model.
Clearly, streaming instability can develop in this model and the percentage of time that this model spends in the unstable regime is substantial -- 45.5\% against 54.5\% in the stable regime. Even if we admit an order of magnitude uncertainty in calculating the Stokes number, streaming instability can still develop in the CSC.  At the same time, the cyan circles calculated for the $\xi=95\%$ model demonstrate that streaming instability cannot  develop in this model.
Finally, we note that the region of streaming instability in Fig.~\ref{stream_cond} is derived assuming an efficient dust settling to the disk midplane, with the dust vertical scale height $H_{\rm d,gr}$ being equal to or smaller than 2\% that of the gas.
The ratio $H_{\rm d,gr}/H_{\rm g}$ can be inferred from  \citet{1995Dubrulle} as
\begin{equation}
    H_{\rm d,gr}=H_{\rm g} \sqrt{{\alpha \over \alpha+St}}.
    \label{scaleH}
\end{equation}
For the typical values of $\rm{St}\sim 0.1$, the required ratio of $H_{\rm d,gr}/H{\rm g}\la0.02$  implies $\alpha\la 10^{-4}$. Such low values of the $\alpha$-parameter can be achieved if the MRI is mostly suppressed in the CSC, which is consistent with an assumed low mass transport rate through the CSC in the $\xi=5\%$ model. We admit that  more accurate numerical simulations extending in the inner sub-au region of the disk are needed to make firmer conclusions. If streaming instability is indeed operational, this
would lead to the formation of solid planetary embryos at orbital positions of $\la$~1.0~au. These protoplanets may further be scattered to larger orbital distances \citep[e.g.][]{2019Marleau} or migrated to smaller distances, gaining in the process gaseous atmospheres according to the core accretion scenario. Disk hydrodynamics simulations coupled to a planetary population synthesis model are needed to further explore this scenario.

We therefore foresee two planet forming mechanisms that may operate concurrently in the disk -- streaming instability in the innermost parts of the disk (working concurrently with the core accretion scenario if solid protoplanetary cores exceed a critical mass) and gravitational fragmentation in the outermost parts of the disk. In the future, we plan to couple our numerical hydrodynamics simulations with a population synthesis model to see if these two mechanisms can explain the observed diversity of exoplantetary characteristics.

\begin{figure}
\includegraphics[width=\linewidth]{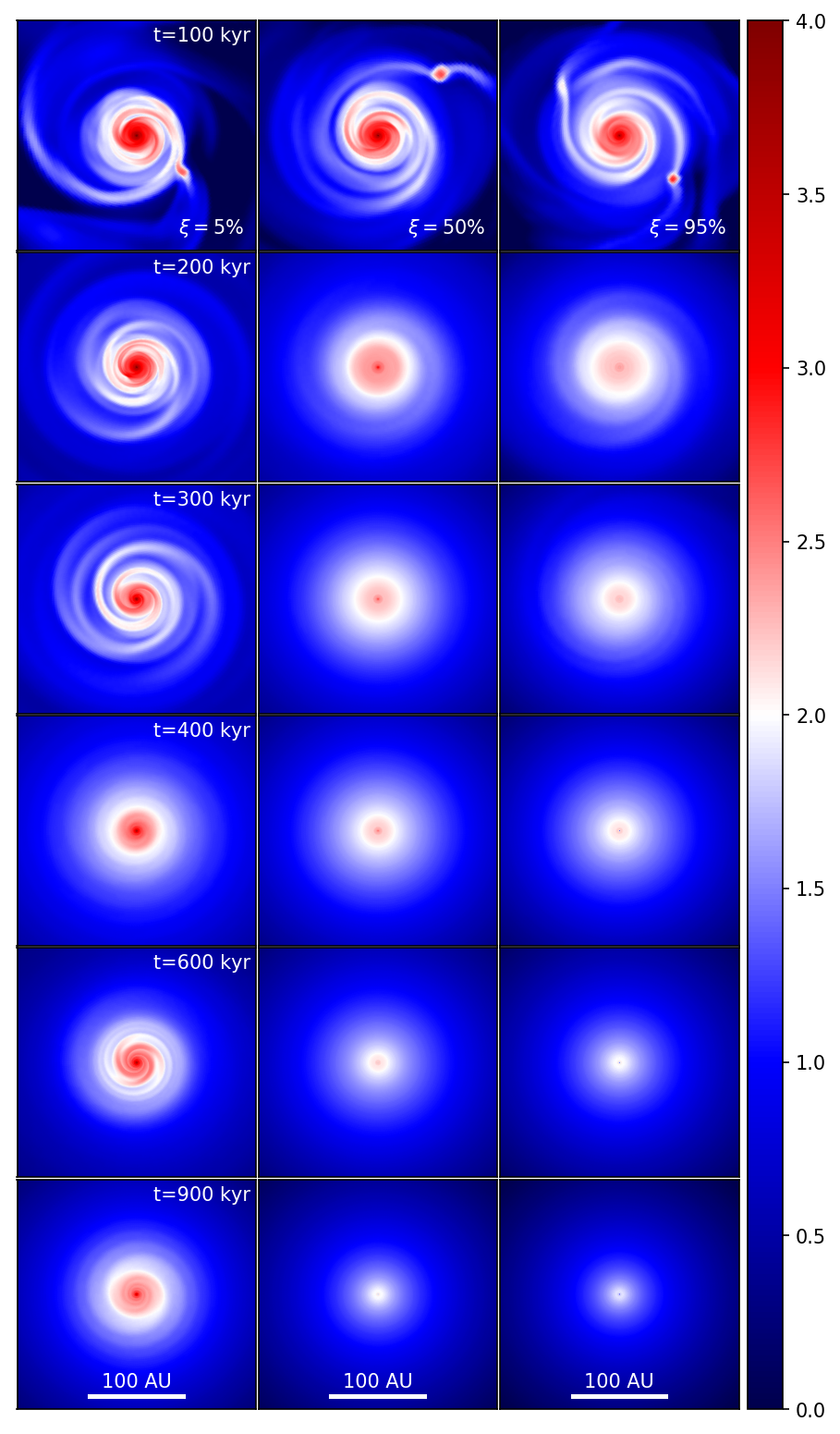}
\caption{Similar to Fig.~\ref{Instab}, but for the twice smaller mass of the prestellar core.}
\label{lowmass2d}
\end{figure}

\subsection{Lower prestellar-core-mass models}

The mass of the prestellar core $M_{\rm core}$ in the models considered in the previous sections was set equal to 1.03 $M_\odot$, i.e., approximately to one solar mass. We only varied the rate of mass transport through the CSC defined by the $\xi$-parameter. It is, however, important to know how sensitive our conclusions may be to the choice of the mass of the prestellar core. In this section, we consider another three models characterized by similar values of the $\xi$-parameters (5\%, 50\%, and 95\%), but having a twice smaller mass of the prestellar core $M_{\rm core}=0.5~M_\odot$.

Figure~\ref{lowmass2d} presents the gas surface densities at several time instances for the three models with a lower mass of the prestellar core. In particular the left, middle, and right columns present the images for the $\xi=5\%$, $\xi=50\%$, and $\xi=95\%$ models, respectively. Clearly, the main evolutionary trends discussed in the context of the one-solar-mass models (see Fig.~\ref{Instab}) hold for the $M_{\rm core}=0.5~M_\odot$ models as well. The disk in the $\xi=5\%$ model demonstrates the presence of gravitational instability to the end of numerical simulations (1.0~Myr), while the disks in the other models become virtually axisymmetric after 0.3~Myr of evolution. All three models, however, are strongly gravitationally unstable in the initial stages of disk formation and evolution ($t\la 100$~kyr).

\begin{figure}
\includegraphics[width=\linewidth]{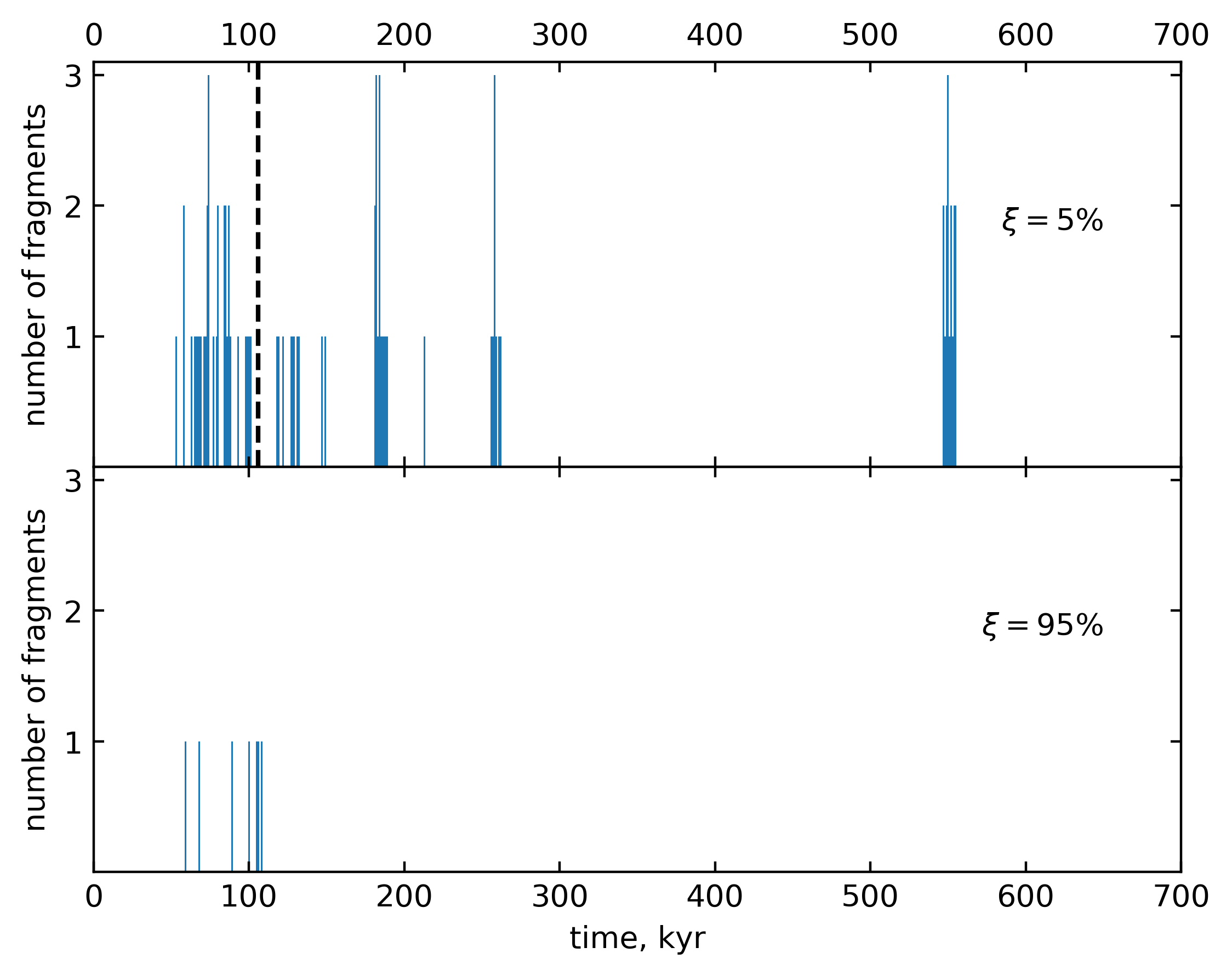}
\caption{Similar to Fig.~\ref{clnum}, but for the twice smaller mass of the prestellar core.}
\label{clnum_lm}
\end{figure}

To further analyze the disk propensity to gravitational fragmentation in the lower-core-mass models, we plot in Figure~\ref{clnum_lm} the number of fragments that exist in the disk as a function of time in the $\xi=5\%$  model (top panel) and $\xi=95\%$ model (bottom panel). A comparison to Fig.~\ref{clnum} showing the number of fragments in the corresponding higher-core-mass models indicates that the lower-core-mass models have systematically smaller numbers of fragments, suggesting less efficient disk fragmentation than in the higher-core-mass case. This is not surprising considering that the disk masses in the lower-core-mass models are factors of several lower. For instance, the disk mass in the $\xi=95\%$ model at the end of numerical  simulations is $0.12~M_\odot$ in the higher-core-mass case and is $0.04~M_\odot$ in the lower-core mass one. What is important is that the episodic fragmentation mode, which was extensively discussed in the context of Fig.~\ref{fig:4} for the higher-core-mass case, also exists in the lower-core-mass regime, albeit less expressed. Indeed, a short episode of disk fragmentation is seen in the $\xi=5\%$ model around $t=550$~kyr, well after the onset of the T Tauri phase shown by the vertical dashed line.

\begin{figure}
\includegraphics[width=\linewidth]{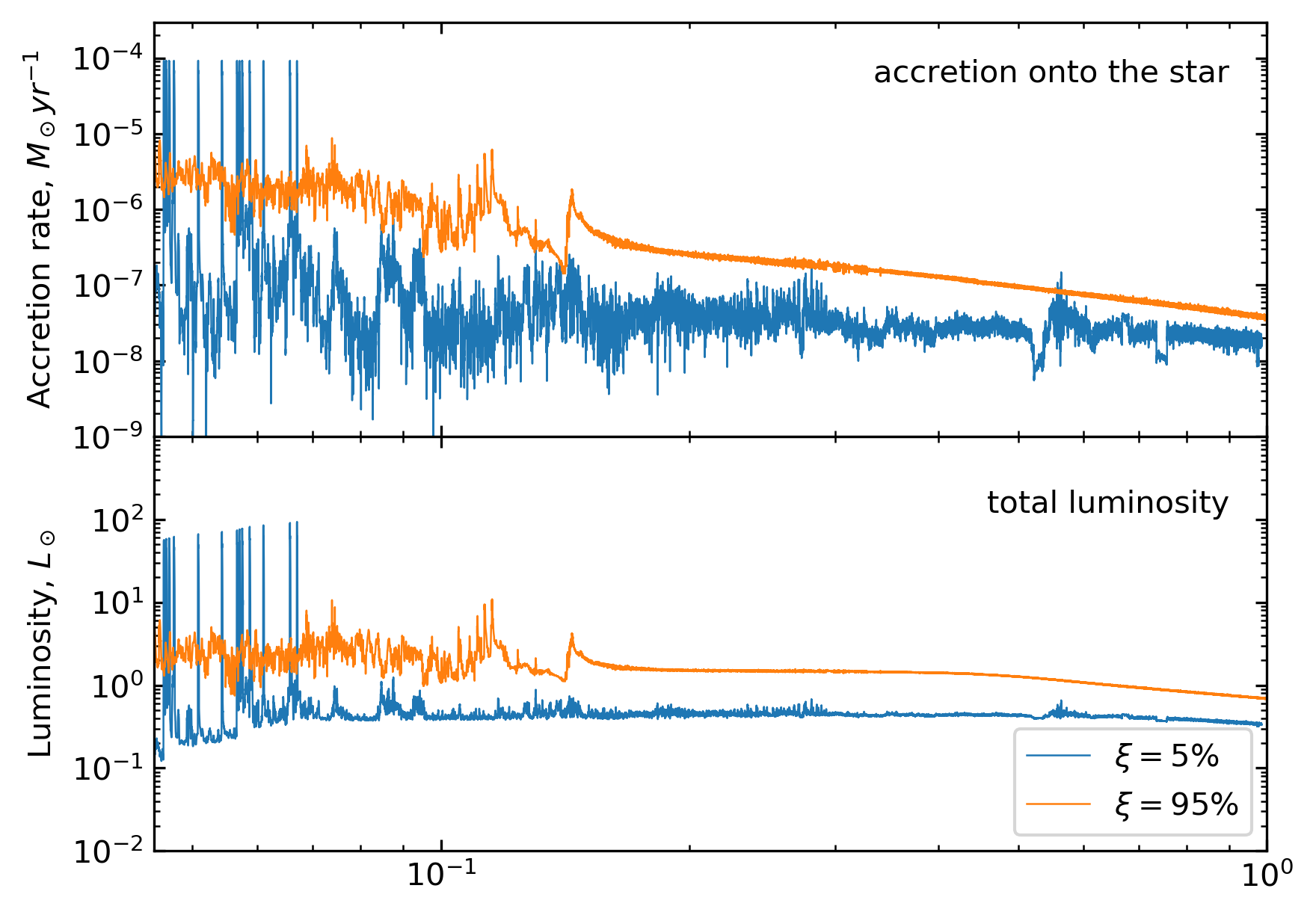}
\caption{Similar to Fig.~\ref{arate01}, but for the twice smaller mass of the prestellar core.}
\label{arates_lm}
\end{figure}

Finally, the mass accretion and total stellar luminosity vs. time for the low-core-mass models are presented in Fig.~\ref{arates_lm}. Clearly, the character of mass accretion and stellar luminosity is qualitatively similar to that discussed  in the context of higher-core-mass models (see Fig.~\ref{arate01}). The $\xi=5\%$ model shows a highly variable accretion with episodic bursts in the early stages of evolution, while the $\xi=95\%$ model is characterized by slowly-varying accretion but systematically higher accretion and stellar luminosity. These qualitative differences in the character of accretion and luminosity between models with different rates of mass transport through the CSC are expected to have important repercussions for the chemical evolution of protoplanetary disks \citep[e.g.,][]{2018MolyarovaAkimkin}.  

\section{Conclusions}
\label{summary}
We studied numerically the global evolution of viscous and self-gravitating circumstellar disks starting from disk formation and ending after about one million years of disk evolution. Such long integration times were made possible thanks to the use of the numerical hydrodynamics code FEOSAD, originally developed by \citet{2018VorobyovAkimkin} and modified in this work to include the effect of back reaction of dust on gas according to the method laid out in \citet{2018Stoyanovskaya}. The turbulent viscosity in the disk is parameterized by a spatially and temporarily constant value of the Shakura-Sunyaev $\alpha$-parameter, set equal to 0.01, which corresponds to the active MRI through the disk extent. Disk self-gravity is calculated via the solution of the Poisson integral.

The inner 1.0 au of the disk is replaced with the central smart cell (CSC), which is a toy model that simulates physical processes that may occur in this region. These processes include  mass transport from the active disk on the nascent star and MRI-triggered accretion bursts.  In this work we focused on one key point -- the dependence of the entire disk evolution on the mass transport rate through the CSC, which is characterized in our models by the $\xi$-parameter. A low value of $\xi=5\%$ implies a slow transport of matter through the CSC on the nascent protostar, while a high value of $\xi=95\%$ means a fast transport. 

Our findings can be summarized as follows.
\begin{itemize}
\item  The evolution of gravitoviscous disks is sensitive to the rate of mass transport through the CSC. More specifically, a slow mass transport rate through the CSC ($\xi=5\%$) leads to the formation of more massive and warmer gaseous disks in comparison to models with a faster mass transport ($\xi=50\%$ and $\xi=95\%$). The maximum size of dust grains is larger in the $\xi=5\%$ model, reaching a decimeter radius. This model also shows an unexpectedly low dust-to-gas mass ratio in the regions between 1~au and 10~au, where it can be factors of several lower than the standard 1:100 value. We argue that this is an effect of gas accumulation in the inner disk, rather than dust depletion.

\item The character of mass accretion on the nascent protostar is notably affected by the rate of mass transport through the CSC. In the $\xi=95\%$ models, the mass accretion rate steadily declines with time with only low-amplitude flickering, while in the $\xi=5\%$ models it is highly  time-variable with episodic bursts caused by the MRI triggering in the CSC. This difference can have important consequences for the dynamical and chemical evolution of the disk. On the one hand, the disk temperature for a steadily declining accretion is on average higher than for a time-variable accretion with episodic bursts. On the other hand, the bursts can raise disk temperature by a factor of several, thus driving chemistry cycles that may be otherwise absent in the disk \citep[e.g.,][]{2018MolyarovaAkimkin}.
We conclude that our results are robust at least for prestellar core masses lying in the $0.5--1.0~M_\odot$ limits.

\item  In the model with slow mass transport through the CSC ($\xi$=5\%) we identified a new mode of episodic disk fragmentation, which occurs in the early T~Tauri phase and alternates with longer quiescent periods of disk stability. 
The masses of fragments formed through disk fragmentation lie in the 2--30~$M_{\rm Jup}$ limits with a maximum at 4--6~$M_{\rm Jup}$.
In general, disks are gravitationally unstable in the embedded phase of evolution, but appear nearly axisymmetric in the T Tauri phase, apart from short-lived episodes of instability, which may be statistically difficult to detect.

\item Grown dust efficiently drifts inward leading to the accumulation of dust in the CSC, so that the highest dust surface densities are found in the inner unresolved 1.0 au of the disk. The dust-to-gas ratios and the Stokes numbers in this region reach the limit when streaming instability is likely to become operational.
We argue that the observed diversity of exoplanets may be explained by two planet-forming mechanisms working concurrently -- streaming instability operating in the innermost disk (together with the core accretion scenario) and gravitational fragmentation operating in the outer disk regions.

\item Our main conclusions hold for protoplanetary disks formed from collapsing prestellar cores with the initial mass lying at least in the 0.5--1.0~$M_\odot$ limits. A study is underway to determine robustness of our results for a wider parameter space.

\end{itemize}

In follow-up studies, we plan to address the effects of a varying $\alpha$-parameter, focusing on the evolution of MRI-suppressed and MRI-dead disks. We will also consider the effects that disk photoevaporation and the varying efficiency of mass ejection with the jets may have on the disk evolution.

\section*{Acknowledgements}
The authors would like to thank the anonymous referee and Beibei Liu for their comments that helped improve the manuscript. This work was supported by the Russian Science Foundation grant 17-12-01168. A.M.S. acknowledges OeAD (Austrian Agency for International Cooperation in Education and Research) for Ernst Mach travel grant for visiting the University of Vienna. V.G.E. acknowledges Swedish Institute for a travel grant allowing to visit the Lund University. The simulations were performed on the Vienna Scientific Cluster. We thank Olga Stoyanovskaya and Dmitri Ionov for fruitful discussions.

\bibliographystyle{aa}
\bibliography{refs}
$$
$$

\end{document}